\documentclass[%
twocolumn,
nofootinbib,
 amsmath,amssymb,
 aps, prd,
]{revtex4-2}
\usepackage{acronym}
\usepackage{xspace}
\usepackage[dvipsnames]{xcolor}
\usepackage[normalem]{ulem}
\usepackage[breaklinks, plainpages=false, colorlinks=true, anchorcolor=cyan, linkcolor=magenta, citecolor=cyan, urlcolor=purple, bookmarks=false]{hyperref}
\usepackage{appendix}
\usepackage{graphicx}
\usepackage{dcolumn}
\usepackage{bm}

\newacro{imbh}[IMBH]{intermediate-mass black hole}
\newacro{bhns}[BHNS]{black hole neutron star}
\newacro{bbh}[BBH]{binary black hole}
\newacro{bh}[BH]{black hole}
\newacro{bns}[BNS]{binary neutron star}
\acrodef{FAR}[FAR]{false alarm rate}
\newacro{bf}[BF]{Bayes' factor}
\newacro{cbc}[CBC]{compact binary coalescence}
\newacro{ce}[CE]{Cosmic Explorer}
\acrodef{SNe}[SNe]{Supernova}
\newacro{da}[DA]{data analysis}
\newacro{et}[ET]{Einstein Telescope}
\newacro{eob}[EOB]{Effective-One-Body}
\newacro{fd}[FD]{frequency domain}
\newacro{gw}[GW]{gravitational-wave}
\newacro{gr}[GR]{general relativity}
\newacro{hm}[HM]{Higher mode}
\newacro{ifo}[IFO]{interferometer}
\newacro{imr}[IMR]{inspiral-merger-ringdown}
\newacro{im}[IM]{inspiral-to-merger}
\newacro{kagra}[KAGRA]{Kamioka Gravitational Wave Detector}
\newacro{ligo}[LIGO]{Laser Interferometer Gravitational-Wave Observatory}
\newacro{lso}[LSO]{Last Stable Orbit}
\newacro{lvc}[LVC]{LIGO-Virgo Collaboration}
\newacro{lvk}[LVK]{LIGO-Virgo-Kagra Collaboration}
\newacro{lo}[LO]{leading order}
\newacro{ns}[NS]{neutron star}
\newacro{nr}[NR]{numerical relativity}
\newacro{pn}[PN]{post-Newtonian}
\newacro{pe}[PE]{parameter estimation}
\newacro{psd}[PSD]{power spectral density}
\newacro{asd}[ASD]{amplitude spectral density}
\acrodef{KN}[KN]{kilonova}
\newacro{xg}[XG]{next-generation}
\newacro{jsd}[JSD]{Jensen Shannon divergence}

\acrodefplural{KN}[KNe]{kilonovae}
\newacro{qc}[QC]{quasi-circular}
\newacro{snr}[SNR]{signal-to-noise ratio}
\acrodef{SNR}[SNR]{signal-to-noise ratio}
\newacro{ng}[NG]{Next Generation}
\newacro{eos}[EoS]{Equation of State}

\newcommand{\nrsur}{\texttt{NRSur7dq4}\xspace}

\newcommand{\gwforge}{\texttt{gwforge}\xspace}

\newcommand{\bilby}{\texttt{Bilby}\xspace}
\newcommand{\dynesty}{\texttt{dynesty}\xspace}

\begin{document}

\preprint{APS/123-QED}

\title{Foreground signals minimally affect inference of high-mass binary black holes in next generation gravitational-wave detectors}
\author{Ish Gupta}
\email{ishgupta@psu.edu}
\affiliation{Institute for Gravitation \& the Cosmos, Department of Physics, The Pennsylvania State University, University Park PA 16802, USA}
\affiliation{Department of Astronomy \& Astrophysics,
    The Pennsylvania State University,
    University Park, PA 16802, USA}

\author{Koustav Chandra}
\email{kbc5795@psu.edu}
\affiliation{Institute for Gravitation \& the Cosmos, Department of Physics, The Pennsylvania State University, University Park PA 16802, USA}
\affiliation{Department of Astronomy \& Astrophysics,
    The Pennsylvania State University,
    University Park, PA 16802, USA}

\author{B. S. Sathyaprakash}
\affiliation{Institute for Gravitation \& the Cosmos, Department of Physics, The Pennsylvania State University, University Park PA 16802, USA}
\affiliation{Department of Astronomy \& Astrophysics,
    The Pennsylvania State University,
    University Park, PA 16802, USA}

\date{\today}

\begin{abstract}
Next-generation gravitational-wave observatories are expected to detect over a thousand compact binary coalescence signals daily, with some lasting from minutes to hours. Consequently, multiple signals will overlap in the time-frequency plane, generating a "foreground noise" that predominantly affects the low-frequency range, where binary neutron star inspiral evolution is gradual. This study investigates the impact of such foreground noise on parameter estimation for short-duration binary black hole signals, particularly those with high detector-frame masses and/or located at large redshifts. Our results show a reduction in detection sensitivity by approximately 25\% when the noise power spectrum deviates by up to 50\% from Gaussian noise due to foreground contamination. Despite this, using standard parameter estimation techniques without subtracting overlapping signals, we find that foreground noise has minimal impact, primarily affecting precision. These findings suggest that even in the presence of substantial foreground noise, global-fit techniques, and/or signal subtraction will not be necessary, as accurate recovery of system parameters is achievable with minimal loss in precision.
   
\end{abstract}

\maketitle

\section{Introduction}\label{sec:introduction}

The \ac{xg} of terrestrial \ac{gw} observatories, such as the \ac{et}~\citep{Hild:2008ng, Punturo:2010zz, Hild:2010id} and \ac{ce}~\citep{Reitze:2019iox, Evans:2021gyd, Evans:2023euw}, are anticipated to begin operations in the mid-2030s. These observatories will offer an order-of-magnitude improvement in sensitivity over current \ac{gw} detectors, with a broader frequency bandwidth that enhances detection capabilities at both low and high frequencies (see Fig.~\ref{fig:design-asd}). Consequently, the detection rate of \acp{cbc} will increase significantly, with observable signals lasting from minutes to hours~\citep{Branchesi:2023mws, Gupta:2023lga, Borhanian:2022czq, Iacovelli:2022bbs, Gupta:2023evt}.

This dramatic improvement will lead to overlapping signals in the detectors' output. The number of overlapping signals depends on both the astrophysical merger rate density and the detectors' extended reach (see Appendix~\ref{sec:newtonian} for an estimate). These signals, particularly those at low frequencies where the inspiral evolution is gradual, will overlap in the time-frequency plane, gradually separating as they approach the merger phase. Current matched-filter searches can uniquely identify such events and determine their merger times with millisecond accuracy~\citep{Relton:2022whr}. Thus, unlike the source confusion expected in LISA, where overlapping white dwarf binary signals may not be uniquely identified~\citep{Umstatter:2005su}, overlapping signals in ground-based \ac{xg} detectors are resolvable and primarily contribute as \textit{foreground noise}~\citep{Wu:2022pyg, Johnson:2024foj, Chandra:2024dhf}.

This foreground noise arises from both detectable and undetectable signals, with the latter typically having a single-detector \ac{snr} of $\lesssim 5$. Although detectable signals can be subtracted from the data, achieving perfect subtraction remains challenging. Significant residuals may persist due to waveform systematics, as demonstrated by \citet{Antonelli:2021vwg}, potentially introducing biases. Moreover, the subtraction process is subject to both statistical and systematic uncertainties, further complicating the analysis.

In addition, \citet{Reali:2022aps} noted that the collection of undetectable signals can affect \ac{pe}, broadening the posterior distributions by distorting the noise \ac{psd} estimates. This arises because the small number of overlapping signals per time-frequency bin does not conform to the central limit theorem, preventing the foreground noise from being modelled as Gaussian. Furthermore, as these signals evolve, the data can no longer be considered wide-sense stationary, violating the assumptions of current noise \ac{psd} estimation methods such as Welch averaging~\citep{Allen:2005fk} and BayesLine~\citep{Cornish:2014kda}. As a result, inaccurate \ac{psd} estimates can degrade the efficacy of \ac{gw} searches and parametric inference. \citet{Pizzati:2021apa} and \citet{Samajdar:2021egv} further showed that overlapping signals with comparable \acp{snr} observed within a short time interval (\(\sim \mathcal{O}(1\,\mathrm{s})\)) can bias posterior estimates, depending on the nature of the source.

In this paper, we investigate the impact of foreground noise on Bayesian posterior estimates when detected signals are not subtracted from the data. Since overlaps predominantly occur at low frequencies, foreground noise primarily affects low-frequency \ac{psd} estimates (see Fig.~\ref{fig:welch-asd}). This is particularly important for high detector-frame mass systems, such as \ac{imbh} binaries or stellar-mass \acp{bbh} at large redshifts~\citep{Jani:2019ffg, Chandra:2023nge, Ng:2020qpk}, as these systems may merge at frequencies below 20 Hz. Therefore, we restrict our focus to these sources.

Prior to the fourth observing run of Advanced LIGO~\citep{LIGOScientific:2014pky} and Virgo~\citep{VIRGO:2014yos}, only a few events with a total detector-frame mass exceeding \(100M_\odot\) have been detected~\citep{Chandra:2021wbw, Nitz:2021zwj, KAGRA:2021vkt}. However, \ac{xg} observatories are expected to detect $\mathcal{O}(1000)$ \ac{imbh} binaries annually, contingent on population assumptions. A subset of these events will be detected with \ac{snr} values on the order of $\mathcal{O}(100)$, allowing source-frame mass constraints with errors as low as $\lesssim 1\%$~\citep{Gupta:2023lga, Reali:2024hqf}.

Binaries originating from Population-III stars are also expected to be detected by \ac{xg} detectors. Current population synthesis studies suggest that the merger rate of these \acp{bbh} may peak at redshift \(z \sim 15\), reaching merger rates of $\mathcal{O}(10^{-1} - 10^{2})/\mathrm{Gpc}^3/\mathrm{yr}$~\citep{Kinugawa:2014zha, Belczynski:2016ieo, Ng:2020qpk, Tanikawa:2021qqi, Ng:2022agi, Santoliquido:2023wzn}. A confident detection of a \ac{bbh} merger at such high redshifts could provide strong evidence of their Population III origins. With total detector-frame masses below $1500 M_\odot$, these systems will merge within the frequency range of \ac{xg} detectors. \citet{Gupta:2023lga} showed that a network of \ac{xg} detectors will not only detect most Population III \ac{bbh} mergers but will also constrain their source-frame masses and redshifts to within $\mathcal{O}(10\%)$ precision.

While all these studies have provided valuable insights, they either neglected foreground noise effects or employed Fisher matrix analyses. Consequently, their parameter precision estimates may not fully capture the limitations of \ac{xg} detectors. This work addresses these gaps by incorporating more realistic data simulations and Bayesian inference techniques. This allows us to more accurately assess the true potential of \ac{xg} observatories in detecting and characterizing intermediate-mass and Population III \acp{bbh}.

Our analysis reveals a reduction in detection sensitivity of  $\sim 25\%$ when the noise power spectrum deviates by approximately 50\% from Gaussian noise levels due to foreground noise. While foreground noise degrades the inferred parameters' precision, potentially limiting the full scientific capability of \ac{xg} detectors, it does not introduce significant biases. Importantly, the true parameter values consistently remain within the 90\% credible intervals. These results indicate that even in scenarios with significant foreground noise, the need for global-fit techniques or signal subtraction may be unnecessary, as the recovery of system parameters remains accurate with only a minimal loss in precision.

The remainder of this paper is organized as follows. In Section~\ref{sec:detector}, we provide an overview of the detector network, detail the likelihood function, and describe the methodology for modelling foreground noise in our analysis. Section~\ref{sec:analysis} outlines the analysis setup and procedures used to generate the results presented in Section~\ref{sec:results}. Finally, Section~\ref{sec:conclusion} presents our conclusions and discusses potential directions for future research.

\section{Detector network, likelihood and foreground noise}\label{sec:detector}

The interplay between foreground noise and detector sensitivity is crucial for understanding how overlapping gravitational wave signals can influence binary parameter inference. 
This section outlines the detector network used in our analysis, the likelihood function employed for signal inference, and the approach to modelling foreground noise.


\subsection{Detector network}

\begin{figure}
    \centering
    \includegraphics[width=0.98\linewidth]{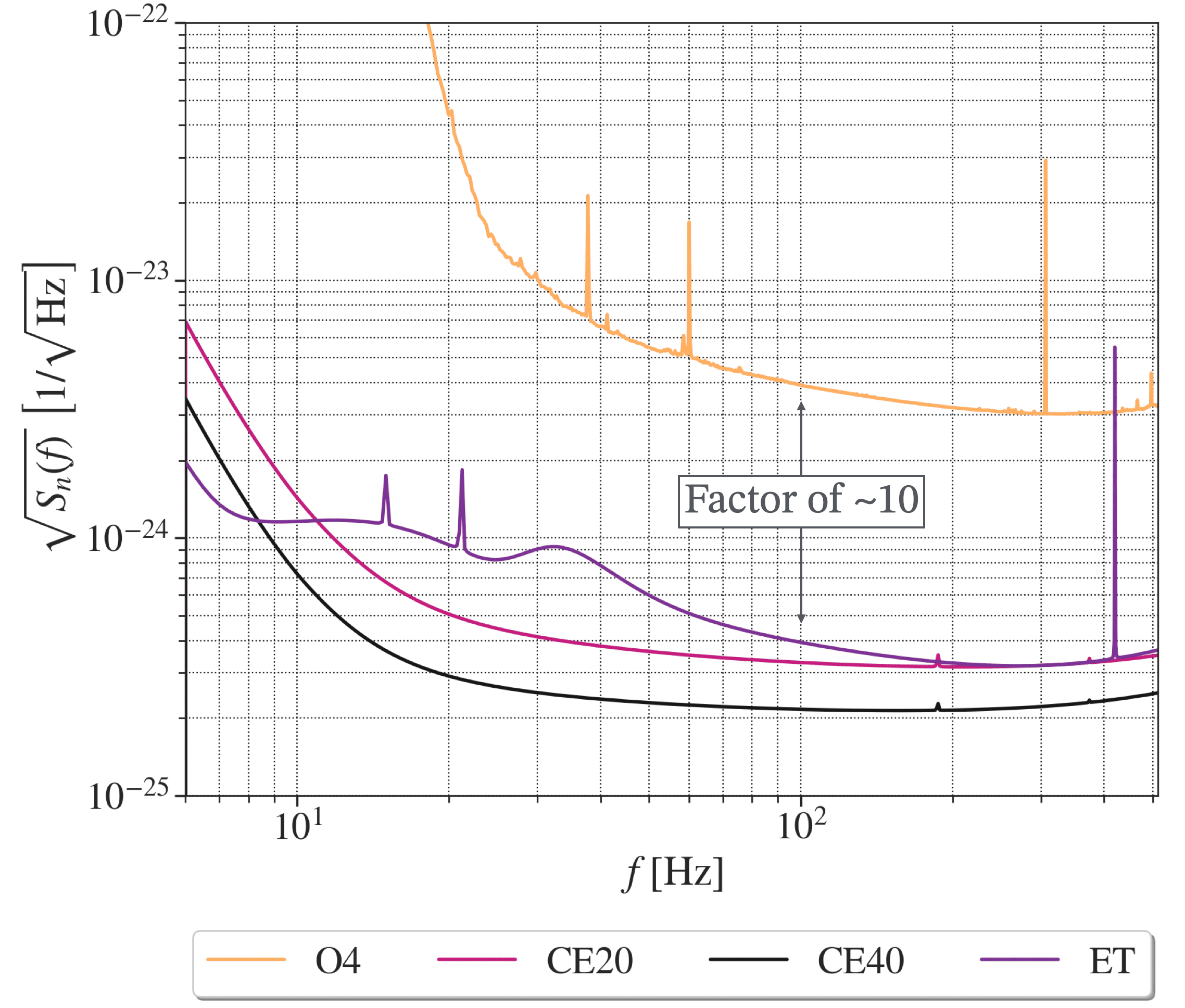}
    \caption{Amplitude spectral densities $\sqrt{S_n(f)}$ of the Advanced LIGO Livingston detector (in O4 run) and \ac{xg} gravitational-wave detectors. The $\sqrt{S_n(f)}$ for Advanced LIGO Livingston is estimated using data surrounding the event GW230529~\citep{LIGOScientific:2024elc, Chandra:2024ila}. The \ac{xg} detectors are expected to achieve sensitivities at least an order of magnitude greater than current detectors, significantly improving the detection capability across a wider frequency range.}
    \label{fig:design-asd}
\end{figure}

We consider a three-detector network composed of a 40 km and a 20 km L-shaped \ac{ce} observatory (henceforth CE40 and CE20), along with a triangle-shaped \ac{et} featuring 10 km arms, as implemented in \gwforge~\citep{Chandra:2024dhf}. Since the exact locations of these detector sites are yet to be finalized, following \citet{MPSACNetwork2024}, we place the \ac{ce} detectors at two sites separated by a distance comparable to the size of the continental US. For \ac{et}, we adopt the potential site in Sardinia. The design noise curves for all these detectors are shown in Fig.~\ref{fig:design-asd}, providing a schematic understanding of how much more sensitive they will be as compared to the Advanced LIGO-Livingston detector in its fourth observing run.

\subsection{Detector data and the likelihood function}
The time-varying output of current \ac{gw} detectors is modelled as the sum of two additive components: the detector noise $\boldsymbol{n}$ and the target signal $\boldsymbol{s}_T(\boldsymbol{\theta})$. For quasi-spherical \ac{bbh} signals, the parameter vector $\boldsymbol{\theta}$ is 15-dimensional and includes the component masses $m_i$, and their spins $\boldsymbol{\chi}_i$, the luminosity distance $D_L$, the sky-location parameters $(\alpha, \delta)$, the binary orientation parameters $(\iota, \phi)$, the polarization angle $\psi$, and the merger time $t_c$.

The noise $\boldsymbol{n}$ is typically assumed to follow a zero-mean, wide-sense stationary, Gaussian process. Under these assumptions, the noise properties are completely characterised by the noise covariance matrix:
\begin{equation}
    \langle |\tilde{n}_k|^2 \rangle = \frac{T}{2} S_n(f_k)~.
\end{equation}
Here $\tilde{n}_k$ is the complex amplitude of the discrete Fourier transform of $\boldsymbol{n}$ at frequency $f_k$, $T$ is the analysis duration, and $S_n(f_k)$ is the noise \ac{psd}.

The objective of modelled signal analysis is to find a template signal $\boldsymbol{h}(\boldsymbol{\theta}')$ that accurately represents the target signal $\boldsymbol{s}_T(\boldsymbol{\theta})$, such that the residual $\boldsymbol{d} - \boldsymbol{h}$ conforms to the noise distribution. This means that the noise model defines the likelihood function. Therefore, for \ac{cbc} analysis, we consider the following likelihood function:
\begin{equation}\label{eq:likelihood}
    \mathfrak{L}\left(\boldsymbol{d} \mid S_n, \boldsymbol{\theta}\right) \propto \exp \left[-\sum_k \frac{2\left|\tilde{d}_k - \tilde{h}_k(\boldsymbol{\theta}')\right|^2}{T S_n(f_k)}\right].
\end{equation}
When analysing a multi-detector signal, we use the joint likelihood, which is the product of individual detector likelihoods, while ensuring that the signal is coherent across the network. This follows from our assumption that non co-located detectors have independent noise properties. Bayesian \ac{pe} routines, such as \bilby \citep{bilby_paper}, aim to estimate the signal parameters by solving:
\begin{equation}
    p(\boldsymbol{\theta} \mid d) \propto \mathcal{L}(d \mid \boldsymbol{\theta}) \pi(\boldsymbol{\theta}),
\end{equation}
for each parameter vector $\boldsymbol{\theta}$ drawn from the joint prior distribution $\pi(\boldsymbol{\theta})$. Thus, the precision of the posterior estimates $p(\boldsymbol{\theta}\mid d)$ is directly influenced by the accuracy of the noise power spectrum $S_n(f)$.

In \ac{xg} detectors, multiple signals will be present at a given instance. As a result, the detector data should be assumed as follows:
\begin{equation}
    \boldsymbol{d} = \boldsymbol{n} + \sum_{k \neq T} \boldsymbol{s}_k(\boldsymbol{\theta}_k) + \boldsymbol{s}_T(\boldsymbol{\theta})~.
\end{equation}
where $\sum_{k \neq T} \boldsymbol{s}_k(\boldsymbol{\theta}_k)$ is the collection of all other signals. This will complicate binary \ac{pe} in two ways. First, the residual $\boldsymbol{d} - \boldsymbol{h}$ is no longer wide-sense stationary as it contains several rapidly-evolving signals. Therefore, it will be incorrect to use Eq.~\eqref{eq:likelihood}. Second, even when estimating the \ac{psd} ``off-source'', meaning computing the \ac{psd} using data segments that exclude the analysis segment that contains $\boldsymbol{s}_T(\boldsymbol{\theta})$, we will still not be measuring the detector's noise properties but will measure the combined effect of the detector noise and the contribution coming from the foreground noise. In other words, we won't be measuring $S_n(f)$ but rather $S_{-s_T}(f)=S_{n + \sum_{k \neq T} \boldsymbol{s}_k}(f)$.

It is possible to overcome both of these problems by jointly inferring the noise power spectrum and $\boldsymbol{\Theta}=\{\boldsymbol{\theta}_k\}$, which is the concatenation of all the binary parameters of all the signals in the analysis segment. However, this is currently computationally prohibitive. Hence, we resort to estimating $S_{-s_T}(f)$ and gauge the affect of this choice on \ac{pe} of short-duration signals.

\subsection{Foreground Noise}\label{sec:foreground}

\begin{figure}
    \centering
    \includegraphics[width=0.98\columnwidth]{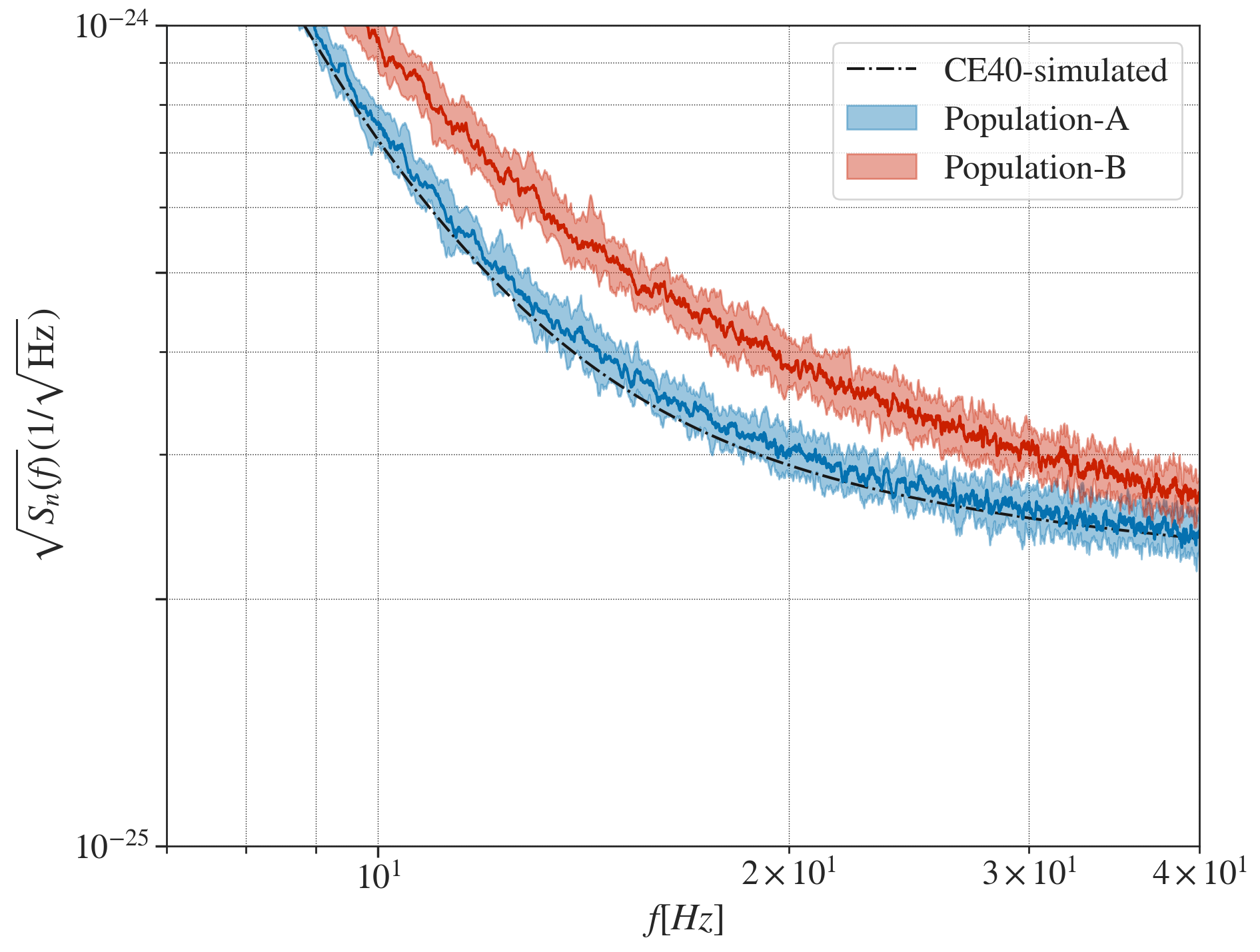}
    \caption{Comparison of Welch estimates for different realizations of 512 seconds of \ac{ce}40 mock data. The black dotted line represents the design \ac{ce}40 sensitivity curve. The red curve indicates the estimate when the noise includes signals from Population A, while the blue curve represents the estimate when the noise includes signals from Population B. Both the red and blue curves include a 90\% confidence band. The comparison demonstrates that depending on the number of signals, the $\sqrt{S_n(f)}$ estimate can deviate from the expected at lower frequencies.}
    \label{fig:welch-asd}
\end{figure}

To demonstrate the impact of foreground noise on \ac{psd} estimates, we use \gwforge to simulate multiple realizations of 8192-seconds long mock \ac{xg} detector data~\citep{Chandra:2024dhf}~\footnote{We take such a long data segment to ensure that the entire length of at least some of \ac{bns} signals are present in a data segment. For most of the signals in the observation band, only a fraction of the signal will be present in a given segment.}. The binary parameter distributions used in the simulations are described in detail in \citet{gwforge-git}. We consider two distinct sets of merger rate densities: Population A, with a \ac{bns} merger rate of $\mathcal{R}_\textrm{BNS}=320/\mathrm{Gpc}^3/\mathrm{yr}$, \ac{bhns} merger rate of $\mathcal{R}_\textrm{BHNS}=45/\mathrm{Gpc}^3/\mathrm{yr}$, and \ac{bbh} merger rate of $\mathcal{R}_\textrm{BBH}=22/\mathrm{Gpc}^3/\mathrm{yr}$, and Population B, with $\mathcal{R}_\textrm{BNS}=1700/\mathrm{Gpc}^3/\mathrm{yr}$, $\mathcal{R}_\textrm{BHNS}=140/\mathrm{Gpc}^3/\mathrm{yr}$, and $\mathcal{R}_\textrm{BBH}=44/\mathrm{Gpc}^3/\mathrm{yr}$. The local merger rates in Population A correspond to the median merger rates inferred using data from the third observing run of Advanced LIGO and Advanced Virgo, whereas the ones in Population B correspond to the upper limit of the inferred rates \citep{KAGRA:2021duu}. The binaries were distributed in redshift following the Madau-Dickinson star-formation rate \citep{Madau:2014bja}. 

The \ac{bns} signals were generated using the quasi-circular \ac{bns} waveform model \texttt{IMRPhenomPv2\_NRTidalv2}~\citep{Dietrich:2019kaq}, while the \ac{bhns} and \ac{bbh} signals were modelled using the quasi-spherical \ac{bbh} approximant \texttt{IMRPhenomXPHM}~\citep{Pratten:2020ceb}. These signals are then added to 8192-second-long realisations of Gaussian noise, coloured by the \ac{xg} detector noise power spectra. This resulted in two distinct datasets: one corresponding to Population A and the other to Population B.

To evaluate the effect of these foreground signals on \ac{psd} estimates, we employed the median-mean Welch method to estimate $S_{-\boldsymbol{s}_T}(f)$ across multiple realizations of both datasets. The Welch method assumes that the data is wide-sense stationary, dividing the discrete time-domain data $d[j]$ into $K$ overlapping segments $d_k[j]$ of length $L$, each windowed by $w[j]$--- in this case, a Hann window function. The discrete Fourier transform $\widetilde{d}_k[i]$ of each windowed segment is computed using an FFT routine, and the periodogram $P_k[i]$ is obtained as follows:
\begin{equation}
    P_k[i] = \frac{1}{U}\left|d_k[i]\right|^2,
\end{equation}
where $U = \sum_{n=0}^{L-1} w[j]^2$. Following \citet{Allen:2005fk}, we then apply mean-median averaging to the periodograms across all segments to obtain the final \ac{psd} estimate, as this method significantly reduces the variance of the spectral estimate due to non-stationarity.

\begin{figure}
    \centering
    \includegraphics[width=0.98\columnwidth]{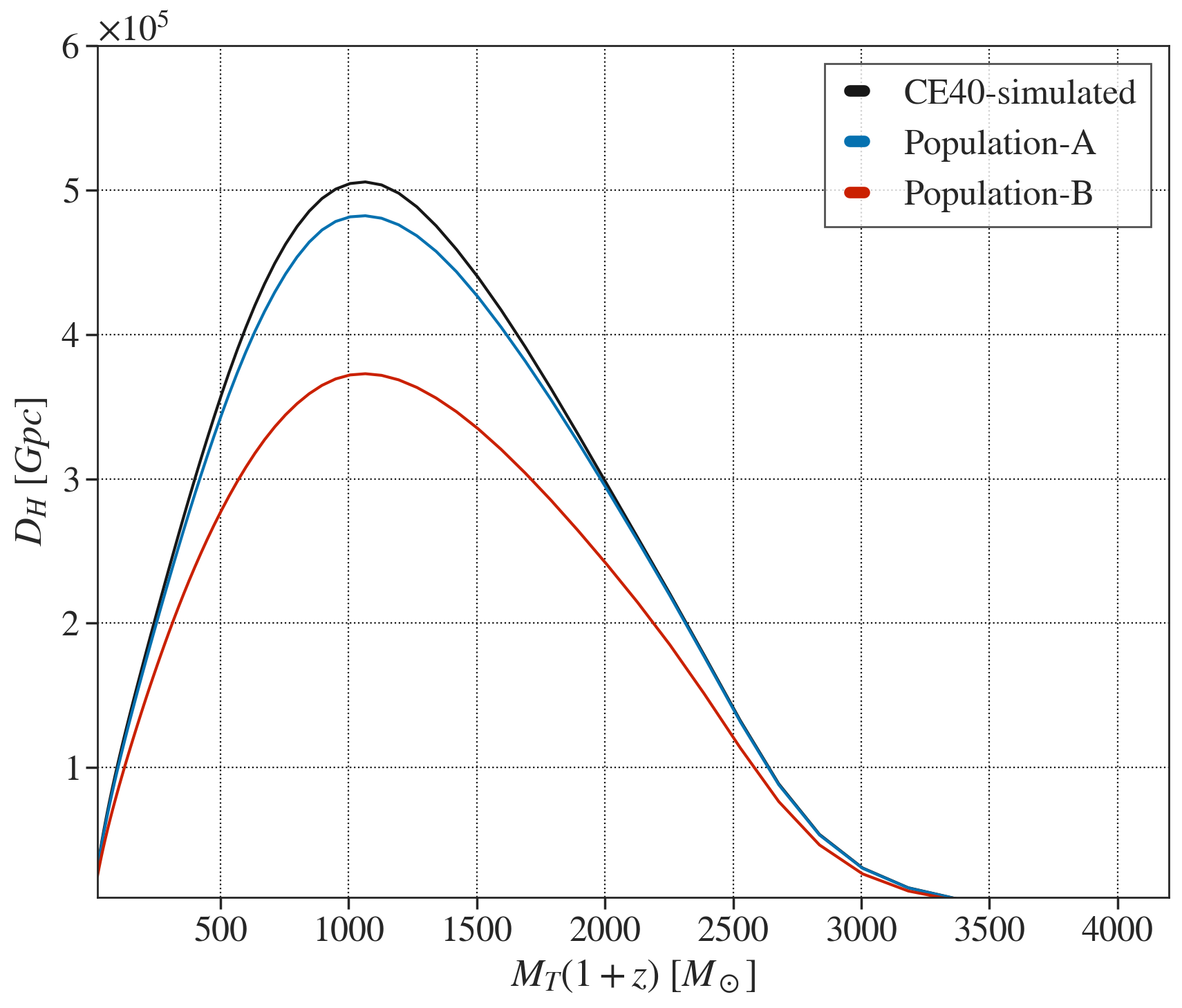}
    \caption{Comparison of horizon distance reach when using the median Welch \ac{asd} estimates. The black dotted line represents the ideal \ac{ce}40 design sensitivity curve. The red curve indicates the estimate when the noise includes signals from population A, while the blue curve represents the estimate when the noise includes signals from population B. The comparison shows that depending on the number of signals, the horizon distance of a \ac{xg} detector can be affected, especially for \acp{bbh} with a high detector frame (redshifted) total mass.}
    \label{fig:horizon}
\end{figure}

Fig.~\ref{fig:welch-asd} displays the simulated and estimated \ac{ce}40 $\sqrt{S_{-\boldsymbol{s}_T}(f)}$, with the latter derived using the Welch method. The estimation was performed by dividing each 512-second realisation of data into 16-second subsegments with 50\% overlap, similar to the approach used in a \textsc{PyCBC} search~\citep{Usman:2015kfa}. Consistent with previous studies~\cite{Wu:2022pyg, Chandra:2024dhf}, we observed that the deviation is primarily concentrated at frequencies lower than 30 Hz, reaching values of $\gtrsim 50\%$ at 15 Hz for the higher merger rate density and CE40 detector. This can reduce the detectability of \acp{bbh} with a high detector frame (redshifted) total mass, as shown in Fig.~\ref{fig:horizon}. In particular, depending on the binary's detector frame mass, the horizon distance~\footnote{The horizon distance, $D_H$, for our case, is the farthest luminosity distance to which we can confidently detect an optimally oriented, located, non-spinning symmetric mass \ac{bbh} assuming an \ac{snr} threshold of 8.} loss to systems with $M_T(1+z) \gtrsim 1000M_\odot$ can be $\gtrsim 30\%$.

\begin{figure*}
    \centering
    \includegraphics[width=0.7\textwidth]{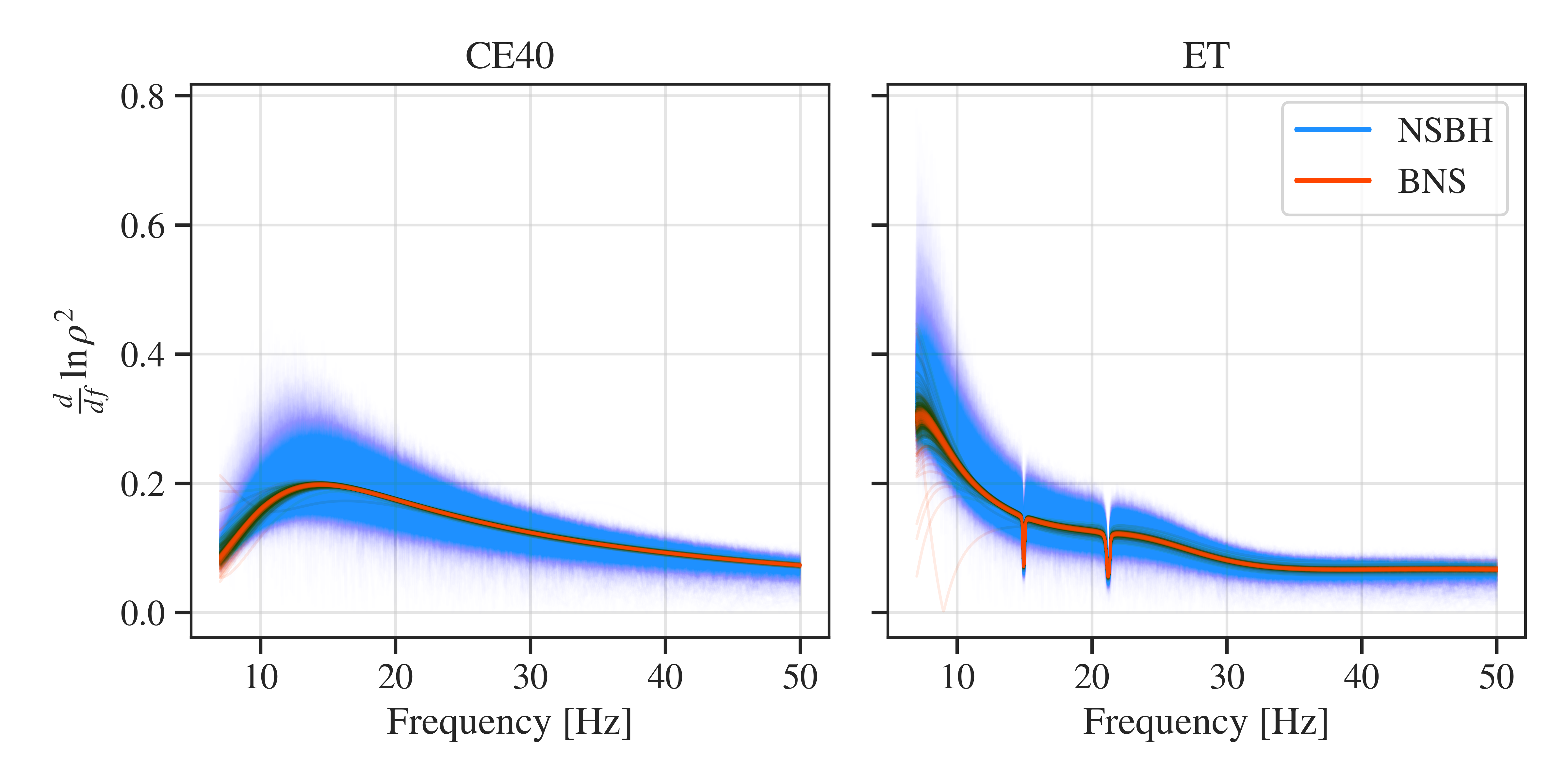}
    \caption{The SNR-squared integrand, normalized by the SNR square, as a frequency function for a simulated population of BNS and NSBH systems in the CE40 detector and in one of the ET detectors.}
    \label{fig:dsnr2_df}
\end{figure*}

To understand the underlying reasons for the observed behaviour, we plot in Fig.~\ref{fig:dsnr2_df} the normalized differential \ac{snr}-squared
\begin{equation}
    \frac{d \ln \rho^2}{df} = \frac{1}{\rho^2} \frac{|\tilde{h}(f)|^2}{S_n(f)}
\end{equation}
for 1000 events picked randomly from the population of \ac{bns} and \ac{bhns} systems using \texttt{gwbench}~\citep{Borhanian:2020ypi}. Here, $S_n(f)$ represents the design noise curve of each detector, and \(\tilde{h}(f)\) is the \ac{gw} signal in the frequency domain. The plot thus shows the fractional contribution to the \ac{snr} as a function of frequency. 

We observe that, for the CE40 detector, the dominant contribution to the  \(d \ln \rho^2 / df\), for both \ac{bns} and \ac{bhns} systems arises in the 10–20 Hz frequency range. The \ac{bhns} systems (blue curves) show a larger variability in \(d \ln \rho^2 / df\) than the \ac{bns} systems (red curves) due to the broader mass distribution of the \ac{bhns} population and the presence of higher-order multipoles in the waveform used to describe them~\citep{Gupta:2024bqn}. In the case of the ET detector, the dominant contribution to \(d \ln \rho^2 / df\) comes from even lower frequencies, as expected, given its greater sensitivity in the sub-10 Hz regime, where the signal spends more time. This enables ET to capture a larger portion of the signal's inspiral.

As the \ac{gw} amplitude decreases $\propto f^{-7/6}$ and the $S_n(f)$ remains relatively flat beyond 20 Hz (up to about 500 Hz), the differential \ac{snr}-squared decreases steadily for both types of systems, thereby having a lesser impact on the Welch \ac{psd} estimate.

\section{Analysis Setup}\label{sec:analysis}

\begin{table*}[]
    \centering
    \begin{tabular}{llll}
        \hline \hline
        Parameter & Symbol & GW150914 & GW190521 \\
        \hline
        Mass ratio & $q$ & 0.98 & 0.95 \\
        Source-frame total mass $(M_{\odot})$ & $M_T$ & 63.05 & 210.51 \\
        Primary spin magnitude & $\chi_1$ & 0.58 & 0.83 \\
        Secondary spin magnitude & $\chi_2$ & 0.43 & 0.96 \\
        Primary tilt (rad) & $\theta_1$ & 2.49 & 1.56 \\
        Secondary tilt (rad) & $\theta_2$ & 0.78 & 1.96 \\
        Azimuthal inter-spin angle (rad) & $\phi_{12}$ & 3.98 & 0.23 \\
        Azimuthal precession cone angle (rad) & $\phi_\mathrm{JL}$ & 0.06 & 0.36 \\
        Effective aligned spin & $\chi_{\textrm{eff }}$ & -0.08 & 0.19 \\
        Effective precessing spin & $\chi_p$ & 0.35 & 0.83 \\
        Azimuth (rad) & $\phi$ & 2.20 & 1.06 \\
        Polarization angle (rad) & $\psi$ & 0.68 & 1.27 \\
        Right ascension (rad) & $\alpha$ & 2.16 & 4.39 \\
        Declination (rad) & $\delta$ & -1.25 & 0.85 \\
        Inclination angle (rad) & $\theta_{J N}$ & 2.55 & 2.09 \\
        \hline \hline
    \end{tabular}
    \caption{Maximum likelihood parameter values for the simulated systems, derived from the GW150914 and GW190521 posteriors as obtained by~\citet{Islam:2023zzj}.}
    \label{tab:injections}
\end{table*}

This section describes the setup used to simulate and analyse \ac{gw} signals in the presence of foreground noise. We begin by outlining the parameters and configurations of the simulated signals, which are based on \ac{bbh} mergers, GW150914 and GW190521. We then detail the inference settings, including the waveform models, prior choices, and sampler configurations used to perform parameter estimation. 

\subsection{Simulated Signals}

We simulate two types of \ac{bbh} signals to explore how foreground noise affects binary \ac{pe}: GW150914-like and GW190521-like, where the former represents a typical Pop-III \ac{bbh} system \citep{Santoliquido:2023wzn}, and the latter a typical \ac{imbh} binary system. All source parameters, except for the luminosity distance \(D_L\), are set to the maximum-likelihood values from the publicly available posterior samples for these events, as reported by \citet{Islam:2023zzj} using the \nrsur waveform model (See Table~\ref{tab:injections}).

The GW150914-like systems are placed at redshifts \(z = 5, 10, 15, 20, 25\), such that the redshifted detector-frame total mass, $M_T(1+z)$, lies between \(378 M_\odot\) and \(1638 M_\odot\). Since larger binary masses reduce the frequency at the merger, effectively limiting the signal's bandwidth, these systems spend most of their evolution in the lower-frequency bands, where the impact of the foreground noise is most significant. Similarly, the GW190521-like systems are placed at redshifts \(z = 1, 2, 4, 6, 8\). These simulated signals are added into multiple noise realisations (with and without foreground noise) to create multiple datasets, each of which is analyzed with the inference settings described below.

\subsection{Inference Settings}

\paragraph{\bf Data and waveform model:}
We conduct full Bayesian \ac{pe} on 16 seconds of simulated data around the target signal using \bilby. Both the signal simulation and \ac{pe} use the \nrsur waveform model~\citep{Varma:2019csw}~\footnote{This ensures that we don't have any bias due to waveform systematics.}, which is trained on quasi-spherical \ac{bbh} simulations from the SXS catalogue~\citep{Boyle:2019kee}, with mass ratios \(q = m_1/m_2 \leq 4\) and dimensionless spins \(\chi_i \leq 0.8\). However, the models work well even when extrapolated outside their training parameter space range up to $q$ = 6.


\paragraph{\bf Priors:}


Consistent with \citet{LIGOScientific:2020ibl}, we adopt uniform priors on spin magnitudes (\(0 \leq \chi_{1,2} \leq 0.99\)) and detector-frame component masses \(m_i(1+z)\), assuming isotropic distributions for spin orientations, sky location, and binary orientation parameters (\(\iota, \phi\)). The priors on the component masses are subject to the following constraints: (i) the mass ratio \(q\) satisfies \(1/6 \leq q \leq 1\); (ii) the total mass \(M_T(1+z)\) exceeds \(100 M_\odot\); and (iii) the chirp mass \(\mathcal{M}_c(1+z)\) lies within the range \(60 M_\odot \leq \mathcal{M}_c(1+z) \leq 400 M_\odot\).

For the luminosity distance, we assume a uniform prior in the source frame, and flat \(\Lambda\)CDM cosmology with Hubble parameter \(H_0 = 67.9 \, \mathrm{km} \, \mathrm{s}^{-1} \, \mathrm{Mpc}^{-1}\) and dark matter density \(\Omega_m = 0.3065\), as in \citet{Romero-Shaw:2020owr, LIGOScientific:2020ibl}, and $D_L$ prior constrained between 1 Gpc and 500 Gpc.

\paragraph{\bf Frequency settings:}
All analyses are performed with a lower frequency cutoff of \(f_\mathrm{low} = 7\) Hz across the detector network. This is also the reference frequency at which all spin posterior distributions are evaluated.

\paragraph{\bf Sampler settings:}
Posterior sampling is performed using the \dynesty sampler in its static configuration, with 1024 live points~\citep{Speagle:2019ivv}. We employ random walks for sampling, with a minimum of 100 and a maximum of 10,000 MCMC steps. The number of random walk steps in each chain is at least 50 times the autocorrelation length (\texttt{nact}). The sampling process continues until the unaccounted evidence fraction, \texttt{dlogz}, reaches 0.1.

\section{Results}\label{sec:results}

This section assesses the impact of foreground noise on the \ac{pe} of GW190521-like and GW150914-like signals. In particular, we show the posteriors for the source-frame component masses, $m_1$ and $m_2$, effective aligned-spin parameter $\chi_\mathrm{eff}$~\footnote{$\chi_\mathrm{eff}$ captures the dominant effect of spin on the binary's inspiral. At a given reference frequency, say $f_\mathrm{ref}$, during the binary's inspiral, it is defined as follows~\citep{Racine:2008qv, Ajith:2009bn, Santamaria:2010yb}:
$$\chi_\mathrm{eff} = \frac{m_1 \chi_1 \mathrm{cos}\theta_1+ m_2 \chi_2 \mathrm{cos}\theta_2}{m_1 + m_2}$$ where $\theta_{k}$ are the component's tilt angle with respect to orbital angular momentum.}, and redshift $z$. These distributions are compared for the cases where the data contains no foreground noise (Gaussian), foreground noise corresponding to Population-A (Pop-A), and foreground noise corresponding to Population-B (Pop-B). 

As \acp{bbh} that form dynamically in active galactic nuclei can produce electromagnetic counterparts (see \citet{Graham:2022xxu} and references therein), accurate localization of these binaries is required for astrophysical and cosmological analyses. Thus, we also present the posterior distributions on the right ascension $(\alpha)$ and the cosine of declination $(\mathrm{cos}\,\delta)$ for all the scenarios.

\begin{figure}
    \centering
    \includegraphics[width=0.95\columnwidth]{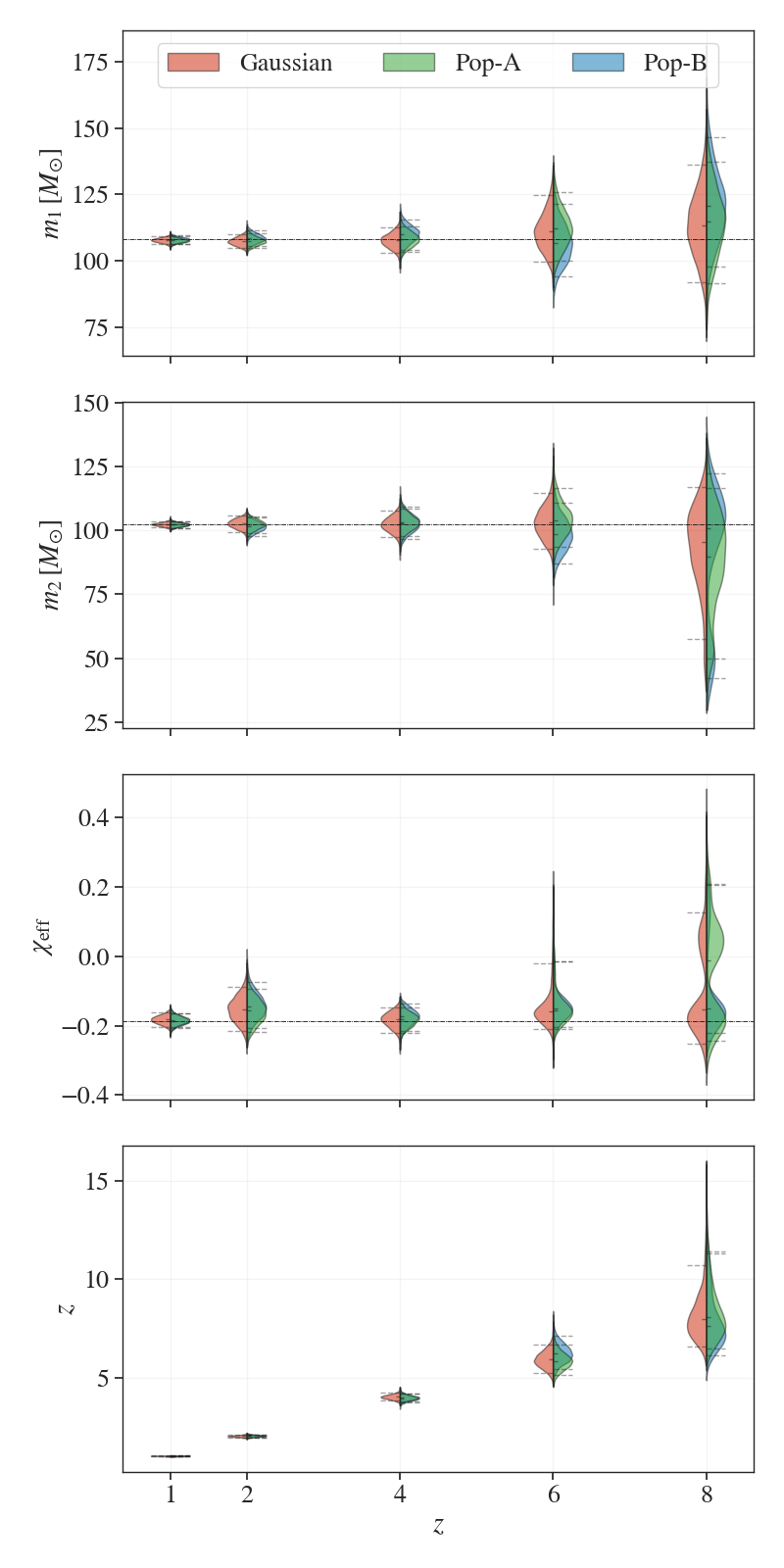}
    \caption{The posterior distribution of the source-frame total mass, mass ratio, and the effective spin parameter for GW190521-like events at different redshifts injected in gaussian noise and foreground noise corresponding to Pop-A and Pop-B. The small solid horizontal line indicates the median of the corresponding distribution, and the dotted horizontal lines show the respective $90\%$ credible interval. The dashed black line shows the true values for $m_1$, $m_2$, and $\chi_{\rm eff}$.}
    \label{fig:190521_pop1vpop2}
\end{figure}

\begin{figure*}
    \centering
    \includegraphics[width=0.8\textwidth]{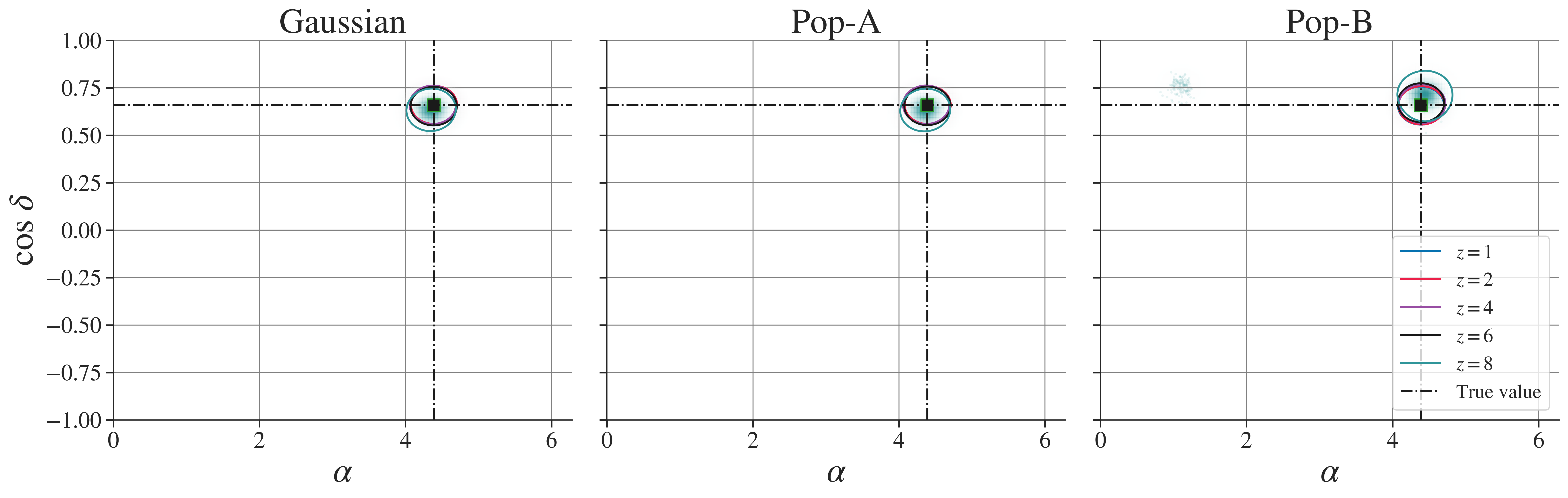}
    \caption{The inferred $90\%$ credible intervals for the right ascension and cosine of declination for GW190521-like \acp{imbh}.}
    \label{fig:190521_sky}
\end{figure*}

\subsection{GW190521-like signals} \label{subsec:190521}
Figure~\ref{fig:190521_pop1vpop2} shows the posteriors for $m_1$, $m_2$, $\chi_{\rm eff}$, and $z$, for GW190521-like events in the three foreground noise scenarios considered in this study. The \acp{snr} for the Gaussian and the Pop-A scenarios are similar for all events, but they reduce by $5\%$-$10\%$ for Pop-B. Across the redshift range explored, we find that the posteriors for the signal parameters for the Gaussian and the Pop-A cases are alike, with the maximum \ac{jsd} value being 0.008 nat, indicating that the precision of \ac{pe} is largely unaffected by differences in the assumed population model.

When comparing the Gaussian and the Pop-B scenarios, we find the posteriors to be broadly consistent, with $\mathrm{\ac{jsd}^{\rm Pop-B}_{\rm Gaussian} \in [0.001,0.08]}$ nat. While the posteriors are expected to be broader for the Pop-B scenario and are seen to deviate slightly from the Gaussian case, the true value lies within the $90\%$ credible interval in all the cases. 

Additionally, we observe that the $\chi_\mathrm{eff}$ posteriors for the $z=2$ system are broader than those for the $z>2$ systems. This broadening exists for the Gaussian case as well. It persists even when the $z=2$ event is injected into different data segments, indicating that foreground noise does not drive the effect. In fact, the effect of foreground noise is minimal for this event, with $\mathrm{\ac{jsd}^{\rm Pop-B}_{\rm Gaussian} = 0.08}$ nat. We investigate the broadening of $\chi_{\rm eff}$ for this event in Appendix~\ref{app:190521-chieff}.

Finally, Figure~\ref{fig:190521_sky} displays the inferred posterior distributions for $\alpha$ and cos$\,\delta$ across different redshifts and foreground noise scenarios. The sky-localization posteriors are consistent across all cases, showing broadening for the Pop-B scenario but no significant deviations. Even in the lowest \ac{snr} case, corresponding to the $z=8$ event with an \ac{snr} $\sim14$ for the Pop-B scenario, the presence of high foreground noise from \acp{cbc} does not bias the sky-localization. This suggests that sky-localization accuracy remains robust, even in scenarios with high merger rates.

\subsection{GW150914-like signals}

\begin{figure}
    \centering
    \includegraphics[width=0.95\columnwidth]{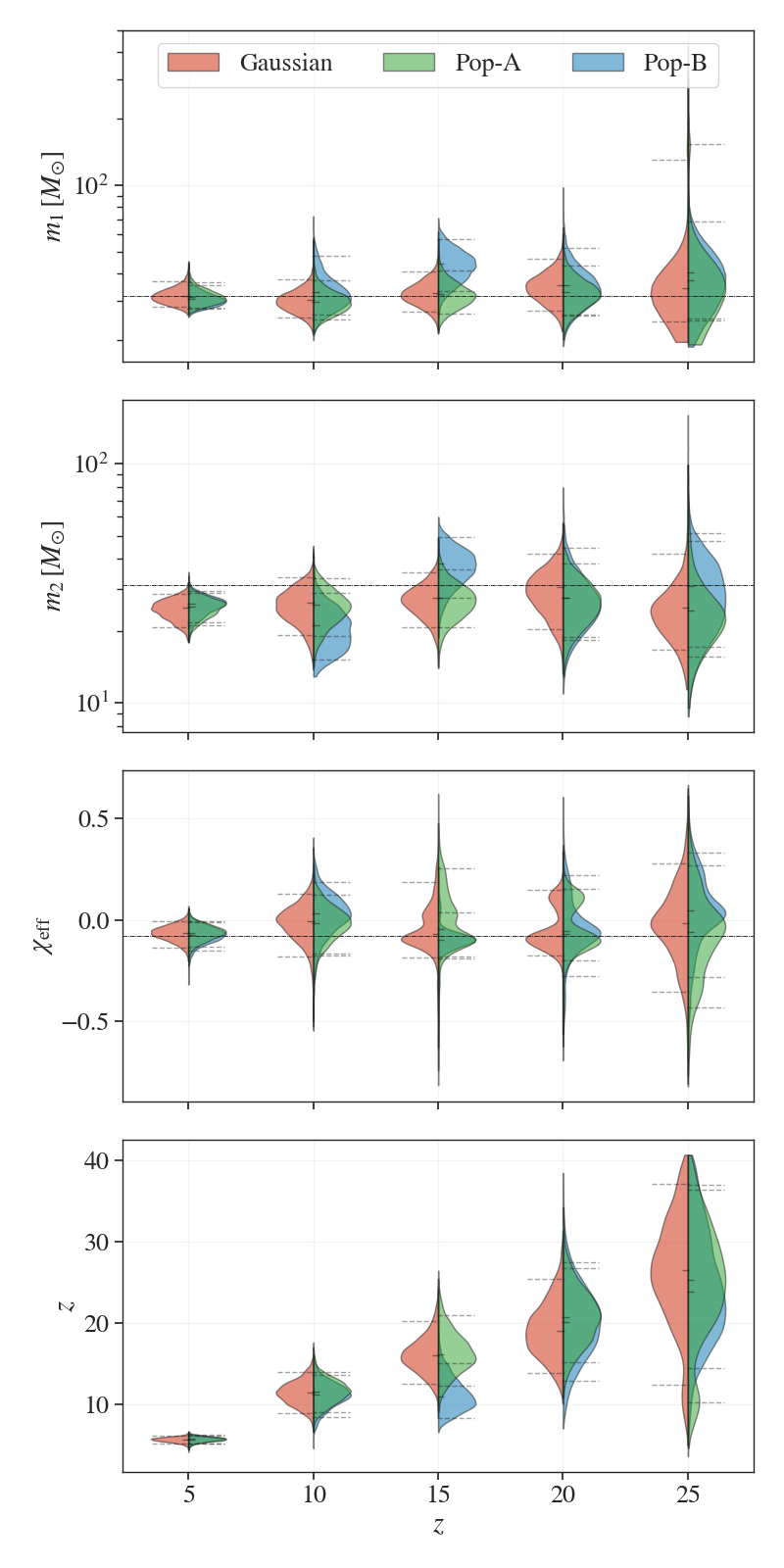}
    \caption{Same as Fig.~\ref{fig:190521_pop1vpop2} but for GW150914-like \acp{bbh} at different redshifts.}
    \label{fig:150914_pop1vpop2}
\end{figure}

\begin{figure*}
    \centering
    \includegraphics[width=0.8\textwidth]{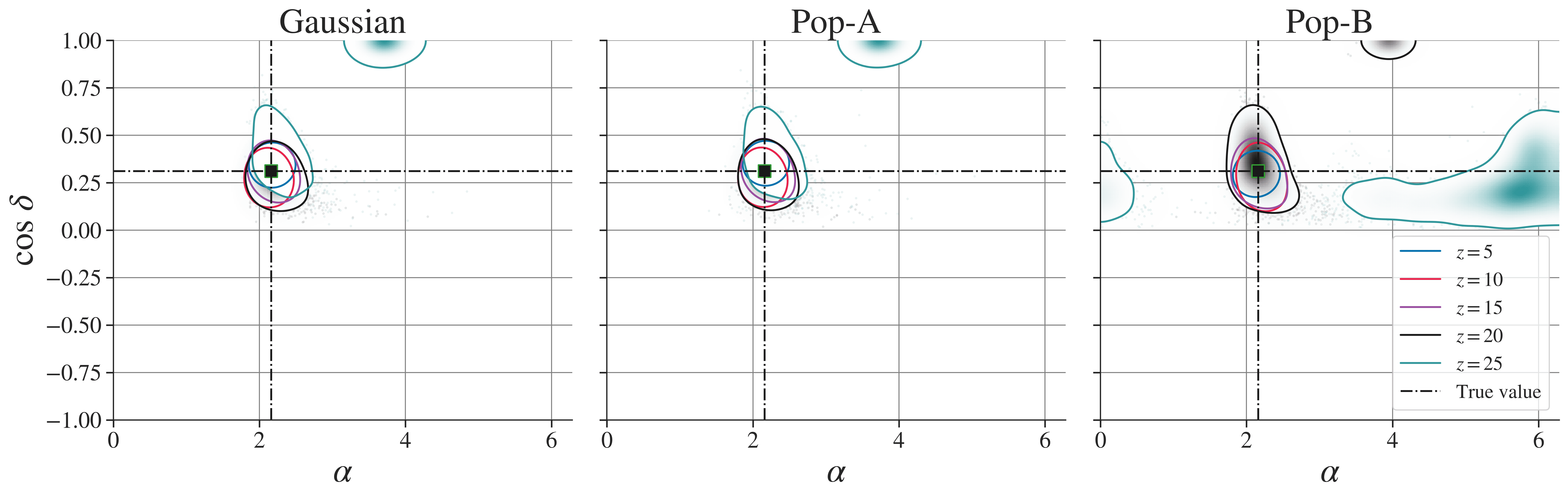}
    \caption{Same as Fig.~\ref{fig:190521_sky} but for GW150914-like \acp{bbh}.}
    \label{fig:150914_sky}
\end{figure*}

Figure~\ref{fig:150914_pop1vpop2} shows the posterior probability distributions for $m_{1,2}$, $\chi_{\rm eff}$, and $z$ for GW150914-like events, representing Pop-III \acp{bbh}. In both the Gaussian and Pop-A scenarios, the \acp{snr} are similar, whereas for Pop-B, they are reduced by approximately $15\%$–$20\%$. Consequently, the posteriors for Pop-B are generally broader. Notably, the inferred parameter distributions in the Gaussian and Pop-A cases are highly consistent, with a maximum \ac{jsd} of only $0.02$ nat across all parameters.

In the Pop-B scenario, the posterior distributions remain broadly consistent with those from the Gaussian case, with a median JSD of 0.03. However, for the $z = 10$ case, we observe a bias in the measurement of $m_2$, where the true value falls outside the $90\%$ credible interval ($\mathrm{JSD}^{\rm Pop-B}_{\rm Gaussian} = 0.1$ nat) but remains within the $3\sigma$ credible region, indicating that the bias is not statistically significant. The bias is also seen to reduce when the same event in injected in different time segments. At $z = 15$, the deviations from the Gaussian and Pop-A cases are more pronounced, with $\mathrm{JSD}^{\rm Pop-B}_{\rm Gaussian} \in [0.01, 0.37]$. While this suggests a larger impact of the foreground noise at this redshift, the true values still lie within the $90\%$ credible intervals. Even under high CBC merger rate conditions, the \ac{pe} for GW150914-like events at high redshifts is not substantially biased, and although the JSD values suggest differences for certain cases, the true values stay within acceptable statistical bounds.


Figure~\ref{fig:150914_sky} shows the inferred distributions of sky location parameters $\alpha$ and $\cos \delta$. The inferred posteriors align well with the true values at all redshifts for both the Gaussian and Pop-A cases. However, for the binary at $z = 25$, an additional mode appears for both, likely due to the influence of Gaussian noise, as the \ac{snr} at this redshift is approximately $13$. In contrast, for Pop-B, the posterior distributions are significantly broader, and the true value falls outside the $90\%$ credible region for the $z = 25$ case.  

Thus, while a foreground composed of \acp{cbc} following median merger rate estimates does not substantially affect source localization, higher merger rate foregrounds degrade both the precision and accuracy of sky localization for GW150914-like events at high redshifts. Although precise localization is essential for electromagnetic follow-up and cosmological studies~\citep{Chen:2020zoq,
Gupta:2022fwd,Chen:2024gdn}, our results suggest that significant deviations from true values primarily occur for low-SNR, extremely distant events, and are therefore not a major concern.

\section{Conclusion}\label{sec:conclusion}

This paper investigates the impact of foreground noise generated by overlapping \ac{cbc} signals in \ac{xg} \ac{gw} observatories. Such noise, originating from numerous in-band signals, renders the detector data non-Gaussian and non-stationary and primarily affects low-frequency sensitivity. 

Unlike previous studies, we neither subtract non-target \ac{gw} signals nor develop new joint signal and noise inference methods. Instead, we employ current techniques to (a) quantify the strength of the foreground noise due to different binary populations, (b) identify the frequencies and conditions where this noise is most impactful, (c) assess its effect on \ac{gw} search sensitivity, and (d) evaluate its influence on binary parameter estimation. 

Our analysis considers three scenarios: (a) Gaussian noise coloured by \ac{xg} design sensitivity curves, (b) Gaussian noise combined with foreground noise from \acp{cbc} at median merger rates (Pop-A), and (c) Gaussian noise with foreground noise from \acp{cbc} population at higher merger rates (Pop-B). We use the Welch method to estimate the \ac{psd}, finding that the \ac{psd} can deviate by up to $50\%$ from Gaussian noise levels for Pop-B at sub-30Hz frequency. This can reduce the detector's horizon reach by $\sim 25\%$.

Given this, we use the state-of-the-art Bayesian inference package \bilby to measure the component source-frame masses $m_k$, effective aligned-spin parameter $\chi_\mathrm{eff}$ and redshift $z$, used interchangeably with luminosity distance $D_L$. We restrict our attention to two different simulated binary populations, namely  GW190521-like \ac{imbh} mergers up to a redshift of 8 and GW150914-like \ac{bbh} mergers (representing mergers of black holes from Pop-III stars) up to a redshift of 25.

We find that for both GW190521-like and GW150914-like systems, the posteriors in the Gaussian and Pop-A cases are highly consistent, with \ac{jsd} values less than 0.01 nat. This demonstrates that foreground noise due to median merger rate estimates does not significantly affect parameter estimation.

However, in the Pop-B case, where the number of signals is higher, we observe a reduction in the SNR by $15\%$–$20\%$ due to the foreground noise. This leads to broader posterior distributions, though biases in parameter estimation remain statistically insignificant (below $3\sigma$). Consequently, we conclude that the foreground noise at high merger rates does not substantially affect the accurate estimation of parameters such as masses, spins, and redshifts. This result is significant as it implies that even in high foreground noise, \ac{xg} detectors can still provide reliable characterization and inference of high-redshift merger properties without subtracting foreground signals.

We also assess the impact of the foreground noise on sky localization, finding that foreground noise does not substantially degrade accuracy, except at low \ac{snr} cases, where mislocalization may occur. This suggests that key science goals, including electromagnetic follow-up and cosmological studies, will not be compromised by foreground noise.

In summary, our findings demonstrate that foreground noise will not significantly affect \ac{pe} for short-duration, high-mass \ac{bbh} systems in \ac{xg} \ac{gw} detectors, even in the presence of high \ac{cbc} merger rates. This result suggests that accurate recovery of system properties is achievable without the need for global fit and/or signal subtraction techniques. Looking ahead, expanding this work to include population-level hierarchical analyses will be crucial for verifying that foreground noise does not bias the inference of hyperparameters that characterize high-redshift populations. However, developing faster and more sophisticated parameter inference algorithms tailored to the intricate noise conditions expected in future observatories will be essential for fully realizing the scientific potential of \ac{xg} \ac{gw} astronomy.

\section*{Acknowledgements}\label{sec:acknowledgements}

The authors thank Tito Dal Canton, Johann Fernandes, and Aditya Vijaykumar for their comments and valuable suggestions. KC would like to thank Jakob Dylan, F. Kapadia and A. J. Hozier-Byrne for their continuous inspiration throughout the development of this work. IG, KC and BSS acknowledge the support through NSF grant numbers PHY-2207638, AST-2307147, PHY-2308886, and PHY-2309064. The authors acknowledge using the Gwave (PSU) cluster for computational/numerical work.

\appendix{}






\section{Number of overlapping signals per frequency bin}
\label{sec:newtonian}
\begin{figure}
    \centering    
    \includegraphics[width=0.9\columnwidth]{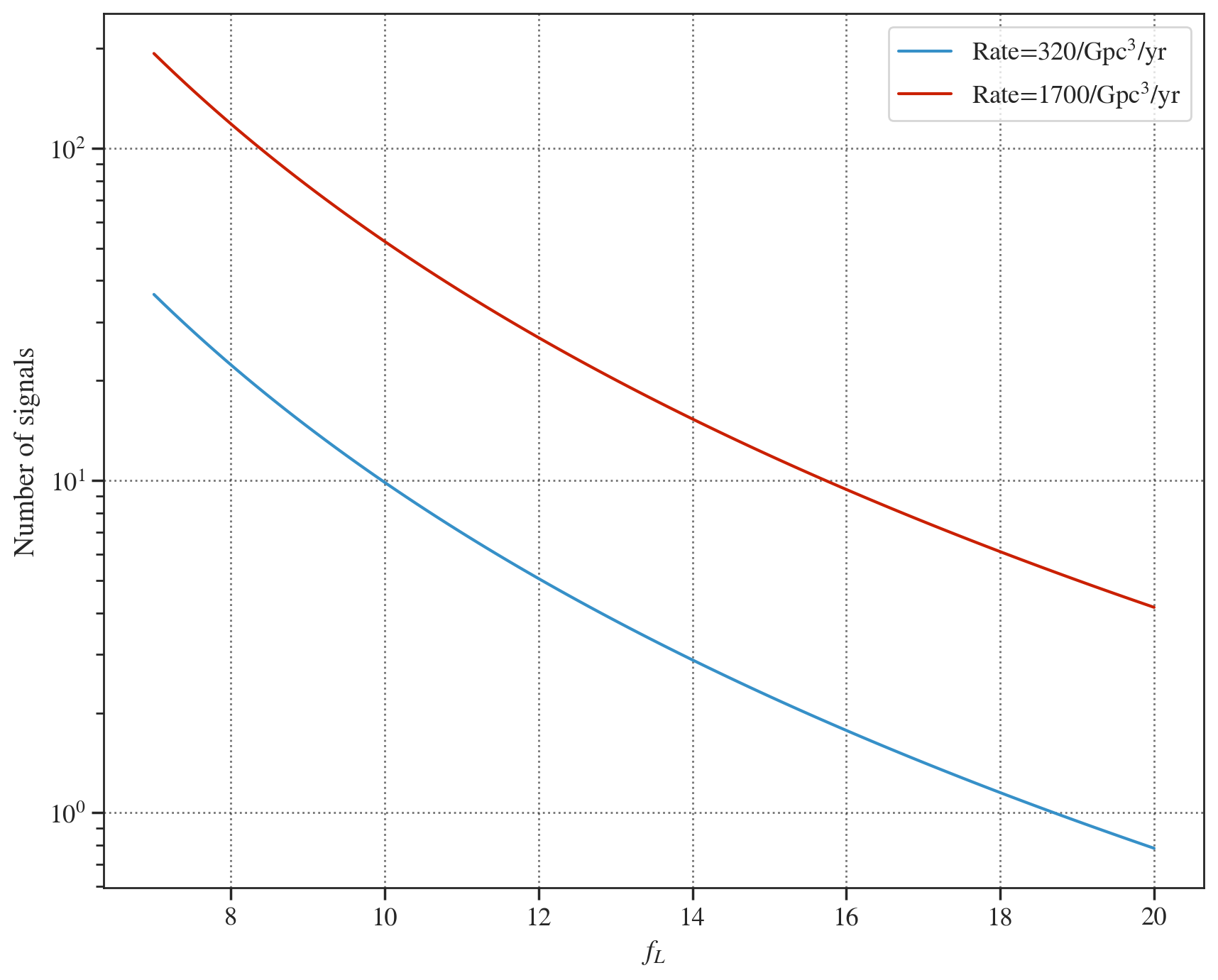}
    \caption{Estimated number of overlapping signals as a function of lower frequency cutoff \( f_L \) for two local merger rate densities.}
    \label{fig:duration}
\end{figure}

\begin{figure}[h]
    \centering
    \includegraphics[width=0.99\columnwidth,trim={0cm 13cm 0cm 0cm}]{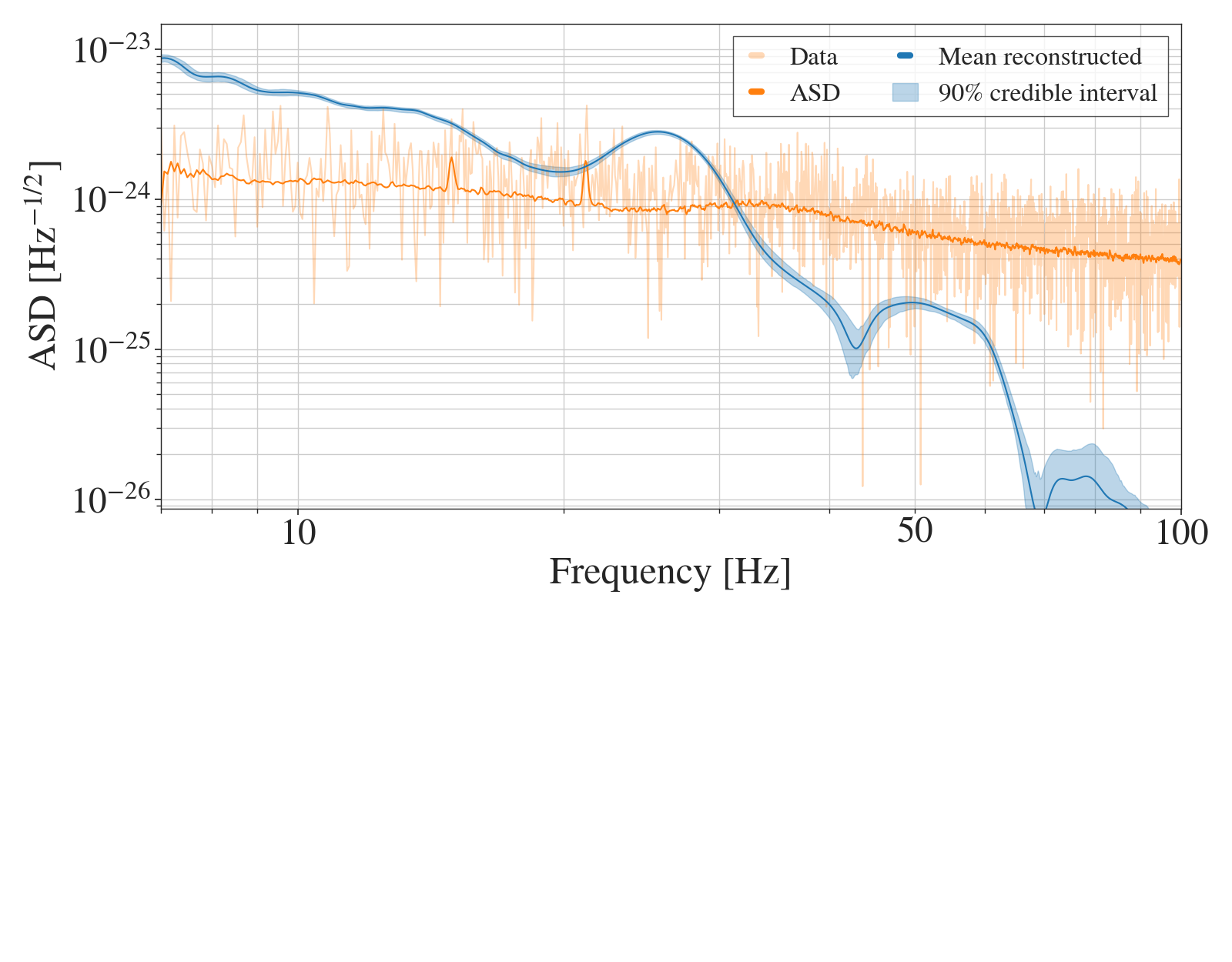}
    \caption{The reconstructed waveform for the GW190521-like \ac{imbh} at z=2, with the \ac{asd} for ET.}
    \label{fig:z_2_waveform}
\end{figure}

\begin{figure}[h]
    \centering    \includegraphics[width=0.99\columnwidth, trim={0cm 1cm 0cm 0cm}]{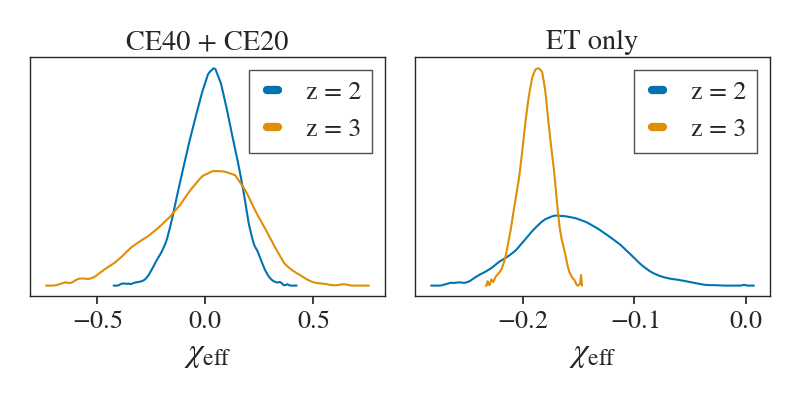}
    \caption{The $\chi_{\rm eff}$ posterior distribution for the $z=2$ and $z=3$ cases in Gaussian noise with CE40 and CE20 detectors, and with only the ET observatory.}
    \label{fig:z_2_chieff}
\end{figure}

We guesstimate the number of overlapping signals per frequency bin as follows.

Consider a binary population with redshifted total mass $M_T (1+z)$ and symmetric mass ratio $\eta = m_1 m_2 / M_T^2$. In the Newtonian limit, the duration of a signal starting from a lower frequency cutoff $f_L$ is given by
\begin{equation}
    \tau_c(f_L) = \frac{5}{256 \eta} \frac{G M_T(1+z)}{c^3} \left( \frac{\pi G M_T(1+z) f_L}{c^3} \right)^{-8/3} .
\end{equation}

If the target signal duration is $t$, then the number of signals in the frequency range $[f_L, f_L + 1/t]$ is determined by
\begin{equation}
    \Delta \tau = \tau_c(f_L) - \tau_c(f_L + 1/t) .
\end{equation}

Assuming a local merger rate density $\mathcal{R}$ (in units of $\mathrm{Gpc}^{-3}\,\mathrm{yr}^{-1}$) and a horizon volume $V_H$ (in Gpc$^3$), the rate of signals per second is
\begin{equation}
    \dot{n} = \frac{\mathcal{R} \times V_H}{365 \times 24 \times 60 \times 60} .
\end{equation}

Thus, the number of signals within $\Delta \tau$ is $\Delta \tau \times \dot{n}$. Figure~\ref{fig:duration} shows the estimated number of overlapping \ac{gw} signals as a function of \( f_L \) for two different local merger rate densities: \( 320 \, \mathrm{Gpc}^{-3} \, \mathrm{yr}^{-1} \) (blue curve) and \( 1700 \, \mathrm{Gpc}^{-3} \, \mathrm{yr}^{-1} \) (red curve). We assume that the system's considered have $M_T(1+z) = 2M_\odot$ and $\eta = 0.25$, thereby giving the upper limit to the number of overlapping signals. As \( f_L \) increases, overlapping signals decrease for both merger rate densities. This trend is expected because higher \( f_L \) values correspond to shorter signal durations, which results in fewer opportunities for signals to overlap within a given frequency bin. 

\section{Investigating $\chi_{\rm eff}$ posterior for GW190521-like event at z=2} 
\label{app:190521-chieff}

In Section~\ref{subsec:190521}, we saw that the effect of foreground noise on \ac{pe} for GW190521-like \acp{imbh} is minimal. However, the posterior distribution obtained for $\chi_{\rm eff}$ for the $z=2$ case was considerably broader than those for $z>2$ events, which is unexpected as the SNR of the latter events is lower than that of $z=2$. As seen in Fig.~\ref{fig:z_2_waveform}, the instrument line in ET detectors intersects the reconstructed waveform just before the merger for the z=2 case. For $z>2$ events, the merger happens below $20$ Hz and is unaffected by this feature. 
To corroborate that the issue occurs due to the noise feature, we re-perform parameter estimation on the $z=2$ and the $z=3$ cases with CE40 and CE20 (no noise feature), and with ET only (with noise feature), in Gaussian noise. This is presented in Fig.~\ref{fig:z_2_chieff}. We see that, unlike Fig.~\ref{fig:190521_pop1vpop2}, the posterior distribution of $\chi_{\rm eff}$ becomes broader, going from $z=2$ to $z=3$, when the events are injected in CE40 and CE20, following our expectation. But when the event is simulated in ET only, the $\chi_{\rm eff}$ is measured more precisely for the $z=3$ case.

\newpage
\bibliography{apssamp}

\begin{thebibliography}{61}%
\makeatletter
\providecommand \@ifxundefined [1]{%
 \@ifx{#1\undefined}
}%
\providecommand \@ifnum [1]{%
 \ifnum #1\expandafter \@firstoftwo
 \else \expandafter \@secondoftwo
 \fi
}%
\providecommand \@ifx [1]{%
 \ifx #1\expandafter \@firstoftwo
 \else \expandafter \@secondoftwo
 \fi
}%
\providecommand \natexlab [1]{#1}%
\providecommand \enquote  [1]{``#1''}%
\providecommand \bibnamefont  [1]{#1}%
\providecommand \bibfnamefont [1]{#1}%
\providecommand \citenamefont [1]{#1}%
\providecommand \href@noop [0]{\@secondoftwo}%
\providecommand \href [0]{\begingroup \@sanitize@url \@href}%
\providecommand \@href[1]{\@@startlink{#1}\@@href}%
\providecommand \@@href[1]{\endgroup#1\@@endlink}%
\providecommand \@sanitize@url [0]{\catcode `\\12\catcode `\$12\catcode `\&12\catcode `\#12\catcode `\^12\catcode `\_12\catcode `\%12\relax}%
\providecommand \@@startlink[1]{}%
\providecommand \@@endlink[0]{}%
\providecommand \url  [0]{\begingroup\@sanitize@url \@url }%
\providecommand \@url [1]{\endgroup\@href {#1}{\urlprefix }}%
\providecommand \urlprefix  [0]{URL }%
\providecommand \Eprint [0]{\href }%
\providecommand \doibase [0]{https://doi.org/}%
\providecommand \selectlanguage [0]{\@gobble}%
\providecommand \bibinfo  [0]{\@secondoftwo}%
\providecommand \bibfield  [0]{\@secondoftwo}%
\providecommand \translation [1]{[#1]}%
\providecommand \BibitemOpen [0]{}%
\providecommand \bibitemStop [0]{}%
\providecommand \bibitemNoStop [0]{.\EOS\space}%
\providecommand \EOS [0]{\spacefactor3000\relax}%
\providecommand \BibitemShut  [1]{\csname bibitem#1\endcsname}%
\let\auto@bib@innerbib\@empty
\bibitem [{\citenamefont {Hild}\ \emph {et~al.}(2008)\citenamefont {Hild}, \citenamefont {Chelkowski},\ and\ \citenamefont {Freise}}]{Hild:2008ng}%
  \BibitemOpen
  \bibfield  {author} {\bibinfo {author} {\bibfnamefont {S.}~\bibnamefont {Hild}}, \bibinfo {author} {\bibfnamefont {S.}~\bibnamefont {Chelkowski}},\ and\ \bibinfo {author} {\bibfnamefont {A.}~\bibnamefont {Freise}},\ }\bibfield  {title} {\bibinfo {title} {{Pushing towards the ET sensitivity using 'conventional' technology}},\ }\Eprint {https://arxiv.org/abs/0810.0604} {arXiv:0810.0604 [gr-qc]}  (\bibinfo {year} {2008})\BibitemShut {NoStop}%
\bibitem [{\citenamefont {Punturo}\ \emph {et~al.}(2010)\citenamefont {Punturo} \emph {et~al.}}]{Punturo:2010zz}%
  \BibitemOpen
  \bibfield  {author} {\bibinfo {author} {\bibfnamefont {M.}~\bibnamefont {Punturo}} \emph {et~al.},\ }\bibfield  {title} {\bibinfo {title} {{The Einstein Telescope: A third-generation gravitational wave observatory}},\ }\href {https://doi.org/10.1088/0264-9381/27/19/194002} {\bibfield  {journal} {\bibinfo  {journal} {Class. Quant. Grav.}\ }\textbf {\bibinfo {volume} {27}},\ \bibinfo {pages} {194002} (\bibinfo {year} {2010})}\BibitemShut {NoStop}%
\bibitem [{\citenamefont {Hild}\ \emph {et~al.}(2011)\citenamefont {Hild} \emph {et~al.}}]{Hild:2010id}%
  \BibitemOpen
  \bibfield  {author} {\bibinfo {author} {\bibfnamefont {S.}~\bibnamefont {Hild}} \emph {et~al.},\ }\bibfield  {title} {\bibinfo {title} {{Sensitivity Studies for Third-Generation Gravitational Wave Observatories}},\ }\href {https://doi.org/10.1088/0264-9381/28/9/094013} {\bibfield  {journal} {\bibinfo  {journal} {Class. Quant. Grav.}\ }\textbf {\bibinfo {volume} {28}},\ \bibinfo {pages} {094013} (\bibinfo {year} {2011})},\ \Eprint {https://arxiv.org/abs/1012.0908} {arXiv:1012.0908 [gr-qc]} \BibitemShut {NoStop}%
\bibitem [{\citenamefont {Reitze}\ \emph {et~al.}(2019)\citenamefont {Reitze} \emph {et~al.}}]{Reitze:2019iox}%
  \BibitemOpen
  \bibfield  {author} {\bibinfo {author} {\bibfnamefont {D.}~\bibnamefont {Reitze}} \emph {et~al.},\ }\bibfield  {title} {\bibinfo {title} {{Cosmic Explorer: The U.S. Contribution to Gravitational-Wave Astronomy beyond LIGO}},\ }\href@noop {} {\bibfield  {journal} {\bibinfo  {journal} {Bull. Am. Astron. Soc.}\ }\textbf {\bibinfo {volume} {51}},\ \bibinfo {pages} {035} (\bibinfo {year} {2019})},\ \Eprint {https://arxiv.org/abs/1907.04833} {arXiv:1907.04833 [astro-ph.IM]} \BibitemShut {NoStop}%
\bibitem [{\citenamefont {Evans}\ \emph {et~al.}(2021)\citenamefont {Evans} \emph {et~al.}}]{Evans:2021gyd}%
  \BibitemOpen
  \bibfield  {author} {\bibinfo {author} {\bibfnamefont {M.}~\bibnamefont {Evans}} \emph {et~al.},\ }\bibfield  {title} {\bibinfo {title} {{A Horizon Study for Cosmic Explorer: Science, Observatories, and Community}},\ }\Eprint {https://arxiv.org/abs/2109.09882} {arXiv:2109.09882 [astro-ph.IM]}  (\bibinfo {year} {2021})\BibitemShut {NoStop}%
\bibitem [{\citenamefont {Evans}\ \emph {et~al.}(2023)\citenamefont {Evans} \emph {et~al.}}]{Evans:2023euw}%
  \BibitemOpen
  \bibfield  {author} {\bibinfo {author} {\bibfnamefont {M.}~\bibnamefont {Evans}} \emph {et~al.},\ }\bibfield  {title} {\bibinfo {title} {{Cosmic Explorer: A Submission to the NSF MPSAC ngGW Subcommittee}},\ }\Eprint {https://arxiv.org/abs/2306.13745} {arXiv:2306.13745 [astro-ph.IM]}  (\bibinfo {year} {2023})\BibitemShut {NoStop}%
\bibitem [{\citenamefont {Branchesi}\ \emph {et~al.}(2023)\citenamefont {Branchesi} \emph {et~al.}}]{Branchesi:2023mws}%
  \BibitemOpen
  \bibfield  {author} {\bibinfo {author} {\bibfnamefont {M.}~\bibnamefont {Branchesi}} \emph {et~al.},\ }\bibfield  {title} {\bibinfo {title} {{Science with the Einstein Telescope: a comparison of different designs}},\ }\href {https://doi.org/10.1088/1475-7516/2023/07/068} {\bibfield  {journal} {\bibinfo  {journal} {JCAP}\ }\textbf {\bibinfo {volume} {07}},\ \bibinfo {pages} {068}},\ \Eprint {https://arxiv.org/abs/2303.15923} {arXiv:2303.15923 [gr-qc]} \BibitemShut {NoStop}%
\bibitem [{\citenamefont {Gupta}\ \emph {et~al.}(2023{\natexlab{a}})\citenamefont {Gupta} \emph {et~al.}}]{Gupta:2023lga}%
  \BibitemOpen
  \bibfield  {author} {\bibinfo {author} {\bibfnamefont {I.}~\bibnamefont {Gupta}} \emph {et~al.},\ }\bibfield  {title} {\bibinfo {title} {{Characterizing Gravitational Wave Detector Networks: From A$^\sharp$ to Cosmic Explorer}},\ }\Eprint {https://arxiv.org/abs/2307.10421} {arXiv:2307.10421 [gr-qc]}  (\bibinfo {year} {2023}{\natexlab{a}})\BibitemShut {NoStop}%
\bibitem [{\citenamefont {Borhanian}\ and\ \citenamefont {Sathyaprakash}(2022)}]{Borhanian:2022czq}%
  \BibitemOpen
  \bibfield  {author} {\bibinfo {author} {\bibfnamefont {S.}~\bibnamefont {Borhanian}}\ and\ \bibinfo {author} {\bibfnamefont {B.~S.}\ \bibnamefont {Sathyaprakash}},\ }\bibfield  {title} {\bibinfo {title} {{Listening to the Universe with Next Generation Ground-Based Gravitational-Wave Detectors}},\ }\Eprint {https://arxiv.org/abs/2202.11048} {arXiv:2202.11048 [gr-qc]}  (\bibinfo {year} {2022})\BibitemShut {NoStop}%
\bibitem [{\citenamefont {Iacovelli}\ \emph {et~al.}(2022)\citenamefont {Iacovelli}, \citenamefont {Mancarella}, \citenamefont {Foffa},\ and\ \citenamefont {Maggiore}}]{Iacovelli:2022bbs}%
  \BibitemOpen
  \bibfield  {author} {\bibinfo {author} {\bibfnamefont {F.}~\bibnamefont {Iacovelli}}, \bibinfo {author} {\bibfnamefont {M.}~\bibnamefont {Mancarella}}, \bibinfo {author} {\bibfnamefont {S.}~\bibnamefont {Foffa}},\ and\ \bibinfo {author} {\bibfnamefont {M.}~\bibnamefont {Maggiore}},\ }\bibfield  {title} {\bibinfo {title} {{Forecasting the Detection Capabilities of Third-generation Gravitational-wave Detectors Using GWFAST}},\ }\href {https://doi.org/10.3847/1538-4357/ac9cd4} {\bibfield  {journal} {\bibinfo  {journal} {Astrophys. J.}\ }\textbf {\bibinfo {volume} {941}},\ \bibinfo {pages} {208} (\bibinfo {year} {2022})},\ \Eprint {https://arxiv.org/abs/2207.02771} {arXiv:2207.02771 [gr-qc]} \BibitemShut {NoStop}%
\bibitem [{\citenamefont {Gupta}\ \emph {et~al.}(2023{\natexlab{b}})\citenamefont {Gupta}, \citenamefont {Borhanian}, \citenamefont {Dhani}, \citenamefont {Chattopadhyay}, \citenamefont {Kashyap}, \citenamefont {Villar},\ and\ \citenamefont {Sathyaprakash}}]{Gupta:2023evt}%
  \BibitemOpen
  \bibfield  {author} {\bibinfo {author} {\bibfnamefont {I.}~\bibnamefont {Gupta}}, \bibinfo {author} {\bibfnamefont {S.}~\bibnamefont {Borhanian}}, \bibinfo {author} {\bibfnamefont {A.}~\bibnamefont {Dhani}}, \bibinfo {author} {\bibfnamefont {D.}~\bibnamefont {Chattopadhyay}}, \bibinfo {author} {\bibfnamefont {R.}~\bibnamefont {Kashyap}}, \bibinfo {author} {\bibfnamefont {V.~A.}\ \bibnamefont {Villar}},\ and\ \bibinfo {author} {\bibfnamefont {B.~S.}\ \bibnamefont {Sathyaprakash}},\ }\bibfield  {title} {\bibinfo {title} {{Neutron star-black hole mergers in next generation gravitational-wave observatories}},\ }\href {https://doi.org/10.1103/PhysRevD.107.124007} {\bibfield  {journal} {\bibinfo  {journal} {Phys. Rev. D}\ }\textbf {\bibinfo {volume} {107}},\ \bibinfo {pages} {124007} (\bibinfo {year} {2023}{\natexlab{b}})},\ \Eprint {https://arxiv.org/abs/2301.08763} {arXiv:2301.08763 [gr-qc]} \BibitemShut {NoStop}%
\bibitem [{\citenamefont {Relton}\ \emph {et~al.}(2022)\citenamefont {Relton}, \citenamefont {Virtuoso}, \citenamefont {Bini}, \citenamefont {Raymond}, \citenamefont {Harry}, \citenamefont {Drago}, \citenamefont {Lazzaro}, \citenamefont {Miani},\ and\ \citenamefont {Tiwari}}]{Relton:2022whr}%
  \BibitemOpen
  \bibfield  {author} {\bibinfo {author} {\bibfnamefont {P.}~\bibnamefont {Relton}}, \bibinfo {author} {\bibfnamefont {A.}~\bibnamefont {Virtuoso}}, \bibinfo {author} {\bibfnamefont {S.}~\bibnamefont {Bini}}, \bibinfo {author} {\bibfnamefont {V.}~\bibnamefont {Raymond}}, \bibinfo {author} {\bibfnamefont {I.}~\bibnamefont {Harry}}, \bibinfo {author} {\bibfnamefont {M.}~\bibnamefont {Drago}}, \bibinfo {author} {\bibfnamefont {C.}~\bibnamefont {Lazzaro}}, \bibinfo {author} {\bibfnamefont {A.}~\bibnamefont {Miani}},\ and\ \bibinfo {author} {\bibfnamefont {S.}~\bibnamefont {Tiwari}},\ }\bibfield  {title} {\bibinfo {title} {{Addressing the challenges of detecting time-overlapping compact binary coalescences}},\ }\href {https://doi.org/10.1103/PhysRevD.106.104045} {\bibfield  {journal} {\bibinfo  {journal} {Phys. Rev. D}\ }\textbf {\bibinfo {volume} {106}},\ \bibinfo {pages} {104045} (\bibinfo {year} {2022})},\ \Eprint {https://arxiv.org/abs/2208.00261} {arXiv:2208.00261 [gr-qc]} \BibitemShut {NoStop}%
\bibitem [{\citenamefont {Umstatter}\ \emph {et~al.}(2005)\citenamefont {Umstatter}, \citenamefont {Christensen}, \citenamefont {Hendry}, \citenamefont {Meyer}, \citenamefont {Simha}, \citenamefont {Veitch}, \citenamefont {Vigeland},\ and\ \citenamefont {Woan}}]{Umstatter:2005su}%
  \BibitemOpen
  \bibfield  {author} {\bibinfo {author} {\bibfnamefont {R.}~\bibnamefont {Umstatter}}, \bibinfo {author} {\bibfnamefont {N.}~\bibnamefont {Christensen}}, \bibinfo {author} {\bibfnamefont {M.}~\bibnamefont {Hendry}}, \bibinfo {author} {\bibfnamefont {R.}~\bibnamefont {Meyer}}, \bibinfo {author} {\bibfnamefont {V.}~\bibnamefont {Simha}}, \bibinfo {author} {\bibfnamefont {J.}~\bibnamefont {Veitch}}, \bibinfo {author} {\bibfnamefont {S.}~\bibnamefont {Vigeland}},\ and\ \bibinfo {author} {\bibfnamefont {G.}~\bibnamefont {Woan}},\ }\bibfield  {title} {\bibinfo {title} {{LISA source confusion: Identification and characterization of signals}},\ }\href {https://doi.org/10.1088/0264-9381/22/18/S04} {\bibfield  {journal} {\bibinfo  {journal} {Class. Quant. Grav.}\ }\textbf {\bibinfo {volume} {22}},\ \bibinfo {pages} {S901} (\bibinfo {year} {2005})},\ \Eprint {https://arxiv.org/abs/gr-qc/0503121} {arXiv:gr-qc/0503121} \BibitemShut {NoStop}%
\bibitem [{\citenamefont {Wu}\ and\ \citenamefont {Nitz}(2023)}]{Wu:2022pyg}%
  \BibitemOpen
  \bibfield  {author} {\bibinfo {author} {\bibfnamefont {S.}~\bibnamefont {Wu}}\ and\ \bibinfo {author} {\bibfnamefont {A.~H.}\ \bibnamefont {Nitz}},\ }\bibfield  {title} {\bibinfo {title} {{Mock data study for next-generation ground-based detectors: The performance loss of matched filtering due to correlated confusion noise}},\ }\href {https://doi.org/10.1103/PhysRevD.107.063022} {\bibfield  {journal} {\bibinfo  {journal} {Phys. Rev. D}\ }\textbf {\bibinfo {volume} {107}},\ \bibinfo {pages} {063022} (\bibinfo {year} {2023})},\ \Eprint {https://arxiv.org/abs/2209.03135} {arXiv:2209.03135 [astro-ph.IM]} \BibitemShut {NoStop}%
\bibitem [{\citenamefont {Johnson}\ \emph {et~al.}(2024)\citenamefont {Johnson}, \citenamefont {Chatziioannou},\ and\ \citenamefont {Farr}}]{Johnson:2024foj}%
  \BibitemOpen
  \bibfield  {author} {\bibinfo {author} {\bibfnamefont {A.~D.}\ \bibnamefont {Johnson}}, \bibinfo {author} {\bibfnamefont {K.}~\bibnamefont {Chatziioannou}},\ and\ \bibinfo {author} {\bibfnamefont {W.~M.}\ \bibnamefont {Farr}},\ }\bibfield  {title} {\bibinfo {title} {{Source confusion from neutron star binaries in ground-based gravitational wave detectors is minimal}},\ }\href {https://doi.org/10.1103/PhysRevD.109.084015} {\bibfield  {journal} {\bibinfo  {journal} {Phys. Rev. D}\ }\textbf {\bibinfo {volume} {109}},\ \bibinfo {pages} {084015} (\bibinfo {year} {2024})},\ \Eprint {https://arxiv.org/abs/2402.06836} {arXiv:2402.06836 [gr-qc]} \BibitemShut {NoStop}%
\bibitem [{\citenamefont {{Chandra}}(2024)}]{Chandra:2024dhf}%
  \BibitemOpen
  \bibfield  {author} {\bibinfo {author} {\bibfnamefont {K.}~\bibnamefont {{Chandra}}},\ }\bibfield  {title} {\bibinfo {title} {{gwforge: A user-friendly package to generate gravitational-wave mock data}},\ }\href {https://doi.org/10.48550/arXiv.2407.21109} {\bibfield  {journal} {\bibinfo  {journal} {arXiv e-prints}\ ,\ \bibinfo {eid} {arXiv:2407.21109}} (\bibinfo {year} {2024})},\ \Eprint {https://arxiv.org/abs/2407.21109} {arXiv:2407.21109 [gr-qc]} \BibitemShut {NoStop}%
\bibitem [{\citenamefont {Antonelli}\ \emph {et~al.}(2021)\citenamefont {Antonelli}, \citenamefont {Burke},\ and\ \citenamefont {Gair}}]{Antonelli:2021vwg}%
  \BibitemOpen
  \bibfield  {author} {\bibinfo {author} {\bibfnamefont {A.}~\bibnamefont {Antonelli}}, \bibinfo {author} {\bibfnamefont {O.}~\bibnamefont {Burke}},\ and\ \bibinfo {author} {\bibfnamefont {J.~R.}\ \bibnamefont {Gair}},\ }\bibfield  {title} {\bibinfo {title} {{Noisy neighbours: inference biases from overlapping gravitational-wave signals}},\ }\href {https://doi.org/10.1093/mnras/stab2358} {\bibfield  {journal} {\bibinfo  {journal} {Mon. Not. Roy. Astron. Soc.}\ }\textbf {\bibinfo {volume} {507}},\ \bibinfo {pages} {5069} (\bibinfo {year} {2021})},\ \Eprint {https://arxiv.org/abs/2104.01897} {arXiv:2104.01897 [gr-qc]} \BibitemShut {NoStop}%
\bibitem [{\citenamefont {Reali}\ \emph {et~al.}(2022)\citenamefont {Reali}, \citenamefont {Antonelli}, \citenamefont {Cotesta}, \citenamefont {Borhanian}, \citenamefont {\c{C}al\i{}\c{s}kan}, \citenamefont {Berti},\ and\ \citenamefont {Sathyaprakash}}]{Reali:2022aps}%
  \BibitemOpen
  \bibfield  {author} {\bibinfo {author} {\bibfnamefont {L.}~\bibnamefont {Reali}}, \bibinfo {author} {\bibfnamefont {A.}~\bibnamefont {Antonelli}}, \bibinfo {author} {\bibfnamefont {R.}~\bibnamefont {Cotesta}}, \bibinfo {author} {\bibfnamefont {S.}~\bibnamefont {Borhanian}}, \bibinfo {author} {\bibfnamefont {M.}~\bibnamefont {\c{C}al\i{}\c{s}kan}}, \bibinfo {author} {\bibfnamefont {E.}~\bibnamefont {Berti}},\ and\ \bibinfo {author} {\bibfnamefont {B.~S.}\ \bibnamefont {Sathyaprakash}},\ }\bibfield  {title} {\bibinfo {title} {{The impact of confusion noise on golden binary neutron-star events in next-generation terrestrial observatories}},\ }\Eprint {https://arxiv.org/abs/2209.13452} {arXiv:2209.13452 [gr-qc]}  (\bibinfo {year} {2022})\BibitemShut {NoStop}%
\bibitem [{\citenamefont {Allen}\ \emph {et~al.}(2012)\citenamefont {Allen}, \citenamefont {Anderson}, \citenamefont {Brady}, \citenamefont {Brown},\ and\ \citenamefont {Creighton}}]{Allen:2005fk}%
  \BibitemOpen
  \bibfield  {author} {\bibinfo {author} {\bibfnamefont {B.}~\bibnamefont {Allen}}, \bibinfo {author} {\bibfnamefont {W.~G.}\ \bibnamefont {Anderson}}, \bibinfo {author} {\bibfnamefont {P.~R.}\ \bibnamefont {Brady}}, \bibinfo {author} {\bibfnamefont {D.~A.}\ \bibnamefont {Brown}},\ and\ \bibinfo {author} {\bibfnamefont {J.~D.~E.}\ \bibnamefont {Creighton}},\ }\bibfield  {title} {\bibinfo {title} {{FINDCHIRP: An Algorithm for detection of gravitational waves from inspiraling compact binaries}},\ }\href {https://doi.org/10.1103/PhysRevD.85.122006} {\bibfield  {journal} {\bibinfo  {journal} {Phys. Rev. D}\ }\textbf {\bibinfo {volume} {85}},\ \bibinfo {pages} {122006} (\bibinfo {year} {2012})},\ \Eprint {https://arxiv.org/abs/gr-qc/0509116} {arXiv:gr-qc/0509116} \BibitemShut {NoStop}%
\bibitem [{\citenamefont {Cornish}\ and\ \citenamefont {Littenberg}(2015)}]{Cornish:2014kda}%
  \BibitemOpen
  \bibfield  {author} {\bibinfo {author} {\bibfnamefont {N.~J.}\ \bibnamefont {Cornish}}\ and\ \bibinfo {author} {\bibfnamefont {T.~B.}\ \bibnamefont {Littenberg}},\ }\bibfield  {title} {\bibinfo {title} {{BayesWave: Bayesian Inference for Gravitational Wave Bursts and Instrument Glitches}},\ }\href {https://doi.org/10.1088/0264-9381/32/13/135012} {\bibfield  {journal} {\bibinfo  {journal} {Class. Quant. Grav.}\ }\textbf {\bibinfo {volume} {32}},\ \bibinfo {pages} {135012} (\bibinfo {year} {2015})},\ \Eprint {https://arxiv.org/abs/1410.3835} {arXiv:1410.3835 [gr-qc]} \BibitemShut {NoStop}%
\bibitem [{\citenamefont {Pizzati}\ \emph {et~al.}(2022)\citenamefont {Pizzati}, \citenamefont {Sachdev}, \citenamefont {Gupta},\ and\ \citenamefont {Sathyaprakash}}]{Pizzati:2021apa}%
  \BibitemOpen
  \bibfield  {author} {\bibinfo {author} {\bibfnamefont {E.}~\bibnamefont {Pizzati}}, \bibinfo {author} {\bibfnamefont {S.}~\bibnamefont {Sachdev}}, \bibinfo {author} {\bibfnamefont {A.}~\bibnamefont {Gupta}},\ and\ \bibinfo {author} {\bibfnamefont {B.}~\bibnamefont {Sathyaprakash}},\ }\bibfield  {title} {\bibinfo {title} {{Toward inference of overlapping gravitational-wave signals}},\ }\href {https://doi.org/10.1103/PhysRevD.105.104016} {\bibfield  {journal} {\bibinfo  {journal} {Phys. Rev. D}\ }\textbf {\bibinfo {volume} {105}},\ \bibinfo {pages} {104016} (\bibinfo {year} {2022})},\ \Eprint {https://arxiv.org/abs/2102.07692} {arXiv:2102.07692 [gr-qc]} \BibitemShut {NoStop}%
\bibitem [{\citenamefont {Samajdar}\ \emph {et~al.}(2021)\citenamefont {Samajdar}, \citenamefont {Janquart}, \citenamefont {Van Den~Broeck},\ and\ \citenamefont {Dietrich}}]{Samajdar:2021egv}%
  \BibitemOpen
  \bibfield  {author} {\bibinfo {author} {\bibfnamefont {A.}~\bibnamefont {Samajdar}}, \bibinfo {author} {\bibfnamefont {J.}~\bibnamefont {Janquart}}, \bibinfo {author} {\bibfnamefont {C.}~\bibnamefont {Van Den~Broeck}},\ and\ \bibinfo {author} {\bibfnamefont {T.}~\bibnamefont {Dietrich}},\ }\bibfield  {title} {\bibinfo {title} {{Biases in parameter estimation from overlapping gravitational-wave signals in the third-generation detector era}},\ }\href {https://doi.org/10.1103/PhysRevD.104.044003} {\bibfield  {journal} {\bibinfo  {journal} {Phys. Rev. D}\ }\textbf {\bibinfo {volume} {104}},\ \bibinfo {pages} {044003} (\bibinfo {year} {2021})},\ \Eprint {https://arxiv.org/abs/2102.07544} {arXiv:2102.07544 [gr-qc]} \BibitemShut {NoStop}%
\bibitem [{\citenamefont {Jani}\ \emph {et~al.}(2019)\citenamefont {Jani}, \citenamefont {Shoemaker},\ and\ \citenamefont {Cutler}}]{Jani:2019ffg}%
  \BibitemOpen
  \bibfield  {author} {\bibinfo {author} {\bibfnamefont {K.}~\bibnamefont {Jani}}, \bibinfo {author} {\bibfnamefont {D.}~\bibnamefont {Shoemaker}},\ and\ \bibinfo {author} {\bibfnamefont {C.}~\bibnamefont {Cutler}},\ }\bibfield  {title} {\bibinfo {title} {{Detectability of Intermediate-Mass Black Holes in Multiband Gravitational Wave Astronomy}},\ }\href {https://doi.org/10.1038/s41550-019-0932-7} {\bibfield  {journal} {\bibinfo  {journal} {Nature Astron.}\ }\textbf {\bibinfo {volume} {4}},\ \bibinfo {pages} {260} (\bibinfo {year} {2019})},\ \Eprint {https://arxiv.org/abs/1908.04985} {arXiv:1908.04985 [gr-qc]} \BibitemShut {NoStop}%
\bibitem [{\citenamefont {Chandra}\ \emph {et~al.}(2024{\natexlab{a}})\citenamefont {Chandra}, \citenamefont {Pai}, \citenamefont {Leong},\ and\ \citenamefont {Calder\'on~Bustillo}}]{Chandra:2023nge}%
  \BibitemOpen
  \bibfield  {author} {\bibinfo {author} {\bibfnamefont {K.}~\bibnamefont {Chandra}}, \bibinfo {author} {\bibfnamefont {A.}~\bibnamefont {Pai}}, \bibinfo {author} {\bibfnamefont {S.~H.~W.}\ \bibnamefont {Leong}},\ and\ \bibinfo {author} {\bibfnamefont {J.}~\bibnamefont {Calder\'on~Bustillo}},\ }\bibfield  {title} {\bibinfo {title} {{Impact of Bayesian priors on the inferred masses of quasicircular intermediate-mass black hole binaries}},\ }\href {https://doi.org/10.1103/PhysRevD.109.104031} {\bibfield  {journal} {\bibinfo  {journal} {Phys. Rev. D}\ }\textbf {\bibinfo {volume} {109}},\ \bibinfo {pages} {104031} (\bibinfo {year} {2024}{\natexlab{a}})},\ \Eprint {https://arxiv.org/abs/2309.01683} {arXiv:2309.01683 [gr-qc]} \BibitemShut {NoStop}%
\bibitem [{\citenamefont {Ng}\ \emph {et~al.}(2021)\citenamefont {Ng}, \citenamefont {Vitale}, \citenamefont {Farr},\ and\ \citenamefont {Rodriguez}}]{Ng:2020qpk}%
  \BibitemOpen
  \bibfield  {author} {\bibinfo {author} {\bibfnamefont {K.~K.~Y.}\ \bibnamefont {Ng}}, \bibinfo {author} {\bibfnamefont {S.}~\bibnamefont {Vitale}}, \bibinfo {author} {\bibfnamefont {W.~M.}\ \bibnamefont {Farr}},\ and\ \bibinfo {author} {\bibfnamefont {C.~L.}\ \bibnamefont {Rodriguez}},\ }\bibfield  {title} {\bibinfo {title} {{Probing multiple populations of compact binaries with third-generation gravitational-wave detectors}},\ }\href {https://doi.org/10.3847/2041-8213/abf8be} {\bibfield  {journal} {\bibinfo  {journal} {Astrophys. J. Lett.}\ }\textbf {\bibinfo {volume} {913}},\ \bibinfo {pages} {L5} (\bibinfo {year} {2021})},\ \Eprint {https://arxiv.org/abs/2012.09876} {arXiv:2012.09876 [astro-ph.CO]} \BibitemShut {NoStop}%
\bibitem [{\citenamefont {Aasi}\ \emph {et~al.}(2015)\citenamefont {Aasi} \emph {et~al.}}]{LIGOScientific:2014pky}%
  \BibitemOpen
  \bibfield  {author} {\bibinfo {author} {\bibfnamefont {J.}~\bibnamefont {Aasi}} \emph {et~al.} (\bibinfo {collaboration} {LIGO Scientific}),\ }\bibfield  {title} {\bibinfo {title} {{Advanced LIGO}},\ }\href {https://doi.org/10.1088/0264-9381/32/7/074001} {\bibfield  {journal} {\bibinfo  {journal} {Class. Quant. Grav.}\ }\textbf {\bibinfo {volume} {32}},\ \bibinfo {pages} {074001} (\bibinfo {year} {2015})},\ \Eprint {https://arxiv.org/abs/1411.4547} {arXiv:1411.4547 [gr-qc]} \BibitemShut {NoStop}%
\bibitem [{\citenamefont {Acernese}\ \emph {et~al.}(2015)\citenamefont {Acernese} \emph {et~al.}}]{VIRGO:2014yos}%
  \BibitemOpen
  \bibfield  {author} {\bibinfo {author} {\bibfnamefont {F.}~\bibnamefont {Acernese}} \emph {et~al.} (\bibinfo {collaboration} {VIRGO}),\ }\bibfield  {title} {\bibinfo {title} {{Advanced Virgo: a second-generation interferometric gravitational wave detector}},\ }\href {https://doi.org/10.1088/0264-9381/32/2/024001} {\bibfield  {journal} {\bibinfo  {journal} {Class. Quant. Grav.}\ }\textbf {\bibinfo {volume} {32}},\ \bibinfo {pages} {024001} (\bibinfo {year} {2015})},\ \Eprint {https://arxiv.org/abs/1408.3978} {arXiv:1408.3978 [gr-qc]} \BibitemShut {NoStop}%
\bibitem [{\citenamefont {Chandra}\ \emph {et~al.}(2021)\citenamefont {Chandra}, \citenamefont {Villa-Ortega}, \citenamefont {Dent}, \citenamefont {McIsaac}, \citenamefont {Pai}, \citenamefont {Harry}, \citenamefont {Davies},\ and\ \citenamefont {Soni}}]{Chandra:2021wbw}%
  \BibitemOpen
  \bibfield  {author} {\bibinfo {author} {\bibfnamefont {K.}~\bibnamefont {Chandra}}, \bibinfo {author} {\bibfnamefont {V.}~\bibnamefont {Villa-Ortega}}, \bibinfo {author} {\bibfnamefont {T.}~\bibnamefont {Dent}}, \bibinfo {author} {\bibfnamefont {C.}~\bibnamefont {McIsaac}}, \bibinfo {author} {\bibfnamefont {A.}~\bibnamefont {Pai}}, \bibinfo {author} {\bibfnamefont {I.~W.}\ \bibnamefont {Harry}}, \bibinfo {author} {\bibfnamefont {G.~S.~C.}\ \bibnamefont {Davies}},\ and\ \bibinfo {author} {\bibfnamefont {K.}~\bibnamefont {Soni}},\ }\bibfield  {title} {\bibinfo {title} {{An optimized PyCBC search for gravitational waves from intermediate-mass black hole mergers}},\ }\href {https://doi.org/10.1103/PhysRevD.104.042004} {\bibfield  {journal} {\bibinfo  {journal} {Phys. Rev. D}\ }\textbf {\bibinfo {volume} {104}},\ \bibinfo {pages} {042004} (\bibinfo {year} {2021})},\ \Eprint {https://arxiv.org/abs/2106.00193} {arXiv:2106.00193 [gr-qc]} \BibitemShut {NoStop}%
\bibitem [{\citenamefont {Nitz}\ \emph {et~al.}(2023)\citenamefont {Nitz}, \citenamefont {Kumar}, \citenamefont {Wang}, \citenamefont {Kastha}, \citenamefont {Wu}, \citenamefont {Sch\"afer}, \citenamefont {Dhurkunde},\ and\ \citenamefont {Capano}}]{Nitz:2021zwj}%
  \BibitemOpen
  \bibfield  {author} {\bibinfo {author} {\bibfnamefont {A.~H.}\ \bibnamefont {Nitz}}, \bibinfo {author} {\bibfnamefont {S.}~\bibnamefont {Kumar}}, \bibinfo {author} {\bibfnamefont {Y.-F.}\ \bibnamefont {Wang}}, \bibinfo {author} {\bibfnamefont {S.}~\bibnamefont {Kastha}}, \bibinfo {author} {\bibfnamefont {S.}~\bibnamefont {Wu}}, \bibinfo {author} {\bibfnamefont {M.}~\bibnamefont {Sch\"afer}}, \bibinfo {author} {\bibfnamefont {R.}~\bibnamefont {Dhurkunde}},\ and\ \bibinfo {author} {\bibfnamefont {C.~D.}\ \bibnamefont {Capano}},\ }\bibfield  {title} {\bibinfo {title} {{4-OGC: Catalog of Gravitational Waves from Compact Binary Mergers}},\ }\href {https://doi.org/10.3847/1538-4357/aca591} {\bibfield  {journal} {\bibinfo  {journal} {Astrophys. J.}\ }\textbf {\bibinfo {volume} {946}},\ \bibinfo {pages} {59} (\bibinfo {year} {2023})},\ \Eprint {https://arxiv.org/abs/2112.06878} {arXiv:2112.06878 [astro-ph.HE]} \BibitemShut {NoStop}%
\bibitem [{\citenamefont {Abbott}\ \emph {et~al.}(2023{\natexlab{a}})\citenamefont {Abbott} \emph {et~al.}}]{KAGRA:2021vkt}%
  \BibitemOpen
  \bibfield  {author} {\bibinfo {author} {\bibfnamefont {R.}~\bibnamefont {Abbott}} \emph {et~al.} (\bibinfo {collaboration} {KAGRA, VIRGO, LIGO Scientific}),\ }\bibfield  {title} {\bibinfo {title} {{GWTC-3: Compact Binary Coalescences Observed by LIGO and Virgo during the Second Part of the Third Observing Run}},\ }\href {https://doi.org/10.1103/PhysRevX.13.041039} {\bibfield  {journal} {\bibinfo  {journal} {Phys. Rev. X}\ }\textbf {\bibinfo {volume} {13}},\ \bibinfo {pages} {041039} (\bibinfo {year} {2023}{\natexlab{a}})},\ \Eprint {https://arxiv.org/abs/2111.03606} {arXiv:2111.03606 [gr-qc]} \BibitemShut {NoStop}%
\bibitem [{\citenamefont {Reali}\ \emph {et~al.}(2024)\citenamefont {Reali}, \citenamefont {Cotesta}, \citenamefont {Antonelli}, \citenamefont {Kritos}, \citenamefont {Strokov},\ and\ \citenamefont {Berti}}]{Reali:2024hqf}%
  \BibitemOpen
  \bibfield  {author} {\bibinfo {author} {\bibfnamefont {L.}~\bibnamefont {Reali}}, \bibinfo {author} {\bibfnamefont {R.}~\bibnamefont {Cotesta}}, \bibinfo {author} {\bibfnamefont {A.}~\bibnamefont {Antonelli}}, \bibinfo {author} {\bibfnamefont {K.}~\bibnamefont {Kritos}}, \bibinfo {author} {\bibfnamefont {V.}~\bibnamefont {Strokov}},\ and\ \bibinfo {author} {\bibfnamefont {E.}~\bibnamefont {Berti}},\ }\href@noop {} {\bibinfo {title} {{Intermediate-mass black hole binary parameter estimation with next-generation ground-based detector networks}}} (\bibinfo {year} {2024}),\ \Eprint {https://arxiv.org/abs/2406.01687} {arXiv:2406.01687 [gr-qc]} \BibitemShut {NoStop}%
\bibitem [{\citenamefont {Kinugawa}\ \emph {et~al.}(2014)\citenamefont {Kinugawa}, \citenamefont {Inayoshi}, \citenamefont {Hotokezaka}, \citenamefont {Nakauchi},\ and\ \citenamefont {Nakamura}}]{Kinugawa:2014zha}%
  \BibitemOpen
  \bibfield  {author} {\bibinfo {author} {\bibfnamefont {T.}~\bibnamefont {Kinugawa}}, \bibinfo {author} {\bibfnamefont {K.}~\bibnamefont {Inayoshi}}, \bibinfo {author} {\bibfnamefont {K.}~\bibnamefont {Hotokezaka}}, \bibinfo {author} {\bibfnamefont {D.}~\bibnamefont {Nakauchi}},\ and\ \bibinfo {author} {\bibfnamefont {T.}~\bibnamefont {Nakamura}},\ }\bibfield  {title} {\bibinfo {title} {{Possible Indirect Confirmation of the Existence of Pop III Massive Stars by Gravitational Wave}},\ }\href {https://doi.org/10.1093/mnras/stu1022} {\bibfield  {journal} {\bibinfo  {journal} {Mon. Not. Roy. Astron. Soc.}\ }\textbf {\bibinfo {volume} {442}},\ \bibinfo {pages} {2963} (\bibinfo {year} {2014})},\ \Eprint {https://arxiv.org/abs/1402.6672} {arXiv:1402.6672 [astro-ph.HE]} \BibitemShut {NoStop}%
\bibitem [{\citenamefont {Belczynski}\ \emph {et~al.}(2017)\citenamefont {Belczynski}, \citenamefont {Ryu}, \citenamefont {Perna}, \citenamefont {Berti}, \citenamefont {Tanaka},\ and\ \citenamefont {Bulik}}]{Belczynski:2016ieo}%
  \BibitemOpen
  \bibfield  {author} {\bibinfo {author} {\bibfnamefont {K.}~\bibnamefont {Belczynski}}, \bibinfo {author} {\bibfnamefont {T.}~\bibnamefont {Ryu}}, \bibinfo {author} {\bibfnamefont {R.}~\bibnamefont {Perna}}, \bibinfo {author} {\bibfnamefont {E.}~\bibnamefont {Berti}}, \bibinfo {author} {\bibfnamefont {T.~L.}\ \bibnamefont {Tanaka}},\ and\ \bibinfo {author} {\bibfnamefont {T.}~\bibnamefont {Bulik}},\ }\bibfield  {title} {\bibinfo {title} {{On the likelihood of detecting gravitational waves from Population III compact object binaries}},\ }\href {https://doi.org/10.1093/mnras/stx1759} {\bibfield  {journal} {\bibinfo  {journal} {Mon. Not. Roy. Astron. Soc.}\ }\textbf {\bibinfo {volume} {471}},\ \bibinfo {pages} {4702} (\bibinfo {year} {2017})},\ \Eprint {https://arxiv.org/abs/1612.01524} {arXiv:1612.01524 [astro-ph.HE]} \BibitemShut {NoStop}%
\bibitem [{\citenamefont {Tanikawa}\ \emph {et~al.}(2022)\citenamefont {Tanikawa}, \citenamefont {Yoshida}, \citenamefont {Kinugawa}, \citenamefont {Trani}, \citenamefont {Hosokawa}, \citenamefont {Susa},\ and\ \citenamefont {Omukai}}]{Tanikawa:2021qqi}%
  \BibitemOpen
  \bibfield  {author} {\bibinfo {author} {\bibfnamefont {A.}~\bibnamefont {Tanikawa}}, \bibinfo {author} {\bibfnamefont {T.}~\bibnamefont {Yoshida}}, \bibinfo {author} {\bibfnamefont {T.}~\bibnamefont {Kinugawa}}, \bibinfo {author} {\bibfnamefont {A.~A.}\ \bibnamefont {Trani}}, \bibinfo {author} {\bibfnamefont {T.}~\bibnamefont {Hosokawa}}, \bibinfo {author} {\bibfnamefont {H.}~\bibnamefont {Susa}},\ and\ \bibinfo {author} {\bibfnamefont {K.}~\bibnamefont {Omukai}},\ }\bibfield  {title} {\bibinfo {title} {{Merger Rate Density of Binary Black Holes through Isolated Population I, II, III and Extremely Metal-poor Binary Star Evolution}},\ }\href {https://doi.org/10.3847/1538-4357/ac4247} {\bibfield  {journal} {\bibinfo  {journal} {Astrophys. J.}\ }\textbf {\bibinfo {volume} {926}},\ \bibinfo {pages} {83} (\bibinfo {year} {2022})},\ \Eprint {https://arxiv.org/abs/2110.10846} {arXiv:2110.10846 [astro-ph.HE]} \BibitemShut {NoStop}%
\bibitem [{\citenamefont {Ng}\ \emph {et~al.}(2022)\citenamefont {Ng}, \citenamefont {Franciolini}, \citenamefont {Berti}, \citenamefont {Pani}, \citenamefont {Riotto},\ and\ \citenamefont {Vitale}}]{Ng:2022agi}%
  \BibitemOpen
  \bibfield  {author} {\bibinfo {author} {\bibfnamefont {K.~K.~Y.}\ \bibnamefont {Ng}}, \bibinfo {author} {\bibfnamefont {G.}~\bibnamefont {Franciolini}}, \bibinfo {author} {\bibfnamefont {E.}~\bibnamefont {Berti}}, \bibinfo {author} {\bibfnamefont {P.}~\bibnamefont {Pani}}, \bibinfo {author} {\bibfnamefont {A.}~\bibnamefont {Riotto}},\ and\ \bibinfo {author} {\bibfnamefont {S.}~\bibnamefont {Vitale}},\ }\bibfield  {title} {\bibinfo {title} {{Constraining High-redshift Stellar-mass Primordial Black Holes with Next-generation Ground-based Gravitational-wave Detectors}},\ }\href {https://doi.org/10.3847/2041-8213/ac7aae} {\bibfield  {journal} {\bibinfo  {journal} {Astrophys. J. Lett.}\ }\textbf {\bibinfo {volume} {933}},\ \bibinfo {pages} {L41} (\bibinfo {year} {2022})},\ \Eprint {https://arxiv.org/abs/2204.11864} {arXiv:2204.11864 [astro-ph.CO]} \BibitemShut {NoStop}%
\bibitem [{\citenamefont {Santoliquido}\ \emph {et~al.}(2023)\citenamefont {Santoliquido}, \citenamefont {Mapelli}, \citenamefont {Iorio}, \citenamefont {Costa}, \citenamefont {Glover}, \citenamefont {Hartwig}, \citenamefont {Klessen},\ and\ \citenamefont {Merli}}]{Santoliquido:2023wzn}%
  \BibitemOpen
  \bibfield  {author} {\bibinfo {author} {\bibfnamefont {F.}~\bibnamefont {Santoliquido}}, \bibinfo {author} {\bibfnamefont {M.}~\bibnamefont {Mapelli}}, \bibinfo {author} {\bibfnamefont {G.}~\bibnamefont {Iorio}}, \bibinfo {author} {\bibfnamefont {G.}~\bibnamefont {Costa}}, \bibinfo {author} {\bibfnamefont {S.~C.~O.}\ \bibnamefont {Glover}}, \bibinfo {author} {\bibfnamefont {T.}~\bibnamefont {Hartwig}}, \bibinfo {author} {\bibfnamefont {R.~S.}\ \bibnamefont {Klessen}},\ and\ \bibinfo {author} {\bibfnamefont {L.}~\bibnamefont {Merli}},\ }\bibfield  {title} {\bibinfo {title} {{Binary black hole mergers from Population III stars: uncertainties from star formation and binary star properties}},\ }\href {https://doi.org/10.1093/mnras/stad1860} {\bibfield  {journal} {\bibinfo  {journal} {Mon. Not. Roy. Astron. Soc.}\ }\textbf {\bibinfo {volume} {524}},\ \bibinfo {pages} {307} (\bibinfo {year} {2023})},\ \bibinfo {note} {[Erratum: Mon.Not.Roy.Astron.Soc. 528, 954--962 (2024)]},\ \Eprint
  {https://arxiv.org/abs/2303.15515} {arXiv:2303.15515 [astro-ph.GA]} \BibitemShut {NoStop}%
\bibitem [{\citenamefont {Abac}\ \emph {et~al.}(2024)\citenamefont {Abac} \emph {et~al.}}]{LIGOScientific:2024elc}%
  \BibitemOpen
  \bibfield  {author} {\bibinfo {author} {\bibfnamefont {A.~G.}\ \bibnamefont {Abac}} \emph {et~al.} (\bibinfo {collaboration} {LIGO Scientific, Virgo,, KAGRA, VIRGO}),\ }\bibfield  {title} {\bibinfo {title} {{Observation of Gravitational Waves from the Coalescence of a 2.5\textendash{}4.5 M $_\odot$ Compact Object and a Neutron Star}},\ }\href {https://doi.org/10.3847/2041-8213/ad5beb} {\bibfield  {journal} {\bibinfo  {journal} {Astrophys. J. Lett.}\ }\textbf {\bibinfo {volume} {970}},\ \bibinfo {pages} {L34} (\bibinfo {year} {2024})},\ \Eprint {https://arxiv.org/abs/2404.04248} {arXiv:2404.04248 [astro-ph.HE]} \BibitemShut {NoStop}%
\bibitem [{\citenamefont {Chandra}\ \emph {et~al.}(2024{\natexlab{b}})\citenamefont {Chandra}, \citenamefont {Gupta}, \citenamefont {Gamba}, \citenamefont {Kashyap}, \citenamefont {Chattopadhyay}, \citenamefont {Gonzalez}, \citenamefont {Bernuzzi},\ and\ \citenamefont {Sathyaprakash}}]{Chandra:2024ila}%
  \BibitemOpen
  \bibfield  {author} {\bibinfo {author} {\bibfnamefont {K.}~\bibnamefont {Chandra}}, \bibinfo {author} {\bibfnamefont {I.}~\bibnamefont {Gupta}}, \bibinfo {author} {\bibfnamefont {R.}~\bibnamefont {Gamba}}, \bibinfo {author} {\bibfnamefont {R.}~\bibnamefont {Kashyap}}, \bibinfo {author} {\bibfnamefont {D.}~\bibnamefont {Chattopadhyay}}, \bibinfo {author} {\bibfnamefont {A.}~\bibnamefont {Gonzalez}}, \bibinfo {author} {\bibfnamefont {S.}~\bibnamefont {Bernuzzi}},\ and\ \bibinfo {author} {\bibfnamefont {B.~S.}\ \bibnamefont {Sathyaprakash}},\ }\bibfield  {title} {\bibinfo {title} {{Everything everywhere all at once: A detailed study of GW230529}},\ }\href@noop {} {\bibfield  {journal} {\bibinfo  {journal} {ArXiv}\ } (\bibinfo {year} {2024}{\natexlab{b}})},\ \Eprint {https://arxiv.org/abs/2405.03841} {arXiv:2405.03841 [astro-ph.HE]} \BibitemShut {NoStop}%
\bibitem [{\citenamefont {Vitale}\ \emph {et~al.}(2024)\citenamefont {Vitale}, \citenamefont {Barsotti}, \citenamefont {Berger}, \citenamefont {Brown}, \citenamefont {Corsi}, \citenamefont {Evans}, \citenamefont {Fairhurst}, \citenamefont {Landry},\ and\ \citenamefont {Sathyaprakash}}]{MPSACNetwork2024}%
  \BibitemOpen
  \bibfield  {author} {\bibinfo {author} {\bibfnamefont {S.}~\bibnamefont {Vitale}}, \bibinfo {author} {\bibfnamefont {L.}~\bibnamefont {Barsotti}}, \bibinfo {author} {\bibfnamefont {E.}~\bibnamefont {Berger}}, \bibinfo {author} {\bibfnamefont {D.}~\bibnamefont {Brown}}, \bibinfo {author} {\bibfnamefont {A.}~\bibnamefont {Corsi}}, \bibinfo {author} {\bibfnamefont {M.}~\bibnamefont {Evans}}, \bibinfo {author} {\bibfnamefont {S.}~\bibnamefont {Fairhurst}}, \bibinfo {author} {\bibfnamefont {M.}~\bibnamefont {Landry}},\ and\ \bibinfo {author} {\bibfnamefont {B.}~\bibnamefont {Sathyaprakash}},\ }\href@noop {} {\bibinfo {title} {Network configurations for mpsac}},\ \bibinfo {howpublished} {\url{https://github.com/cosmic-explorer/mpsac_detector_networks/blob/main/doc/Network_configurations_for_MPSAC.pdf}} (\bibinfo {year} {2024}),\ \bibinfo {note} {accessed: 2024-09-02}\BibitemShut {NoStop}%
\bibitem [{\citenamefont {Ashton}\ \emph {et~al.}(2019)\citenamefont {Ashton} \emph {et~al.}}]{bilby_paper}%
  \BibitemOpen
  \bibfield  {author} {\bibinfo {author} {\bibfnamefont {G.}~\bibnamefont {Ashton}} \emph {et~al.},\ }\bibfield  {title} {\bibinfo {title} {{BILBY: A user-friendly Bayesian inference library for gravitational-wave astronomy}},\ }\href {https://doi.org/10.3847/1538-4365/ab06fc} {\bibfield  {journal} {\bibinfo  {journal} {Astrophys. J. Suppl.}\ }\textbf {\bibinfo {volume} {241}},\ \bibinfo {pages} {27} (\bibinfo {year} {2019})},\ \Eprint {https://arxiv.org/abs/1811.02042} {arXiv:1811.02042 [astro-ph.IM]} \BibitemShut {NoStop}%
\bibitem [{\citenamefont {Chandra}(2024)}]{gwforge-git}%
  \BibitemOpen
  \bibfield  {author} {\bibinfo {author} {\bibfnamefont {K.}~\bibnamefont {Chandra}},\ }\href@noop {} {\bibinfo {title} {Gwforge: A lightweight code to generate mock gravitational wave detector data}},\ \bibinfo {howpublished} {\url{https://github.com/koustavchandra/gwforge}} (\bibinfo {year} {2024}),\ \bibinfo {note} {accessed: 2024-05-14}\BibitemShut {NoStop}%
\bibitem [{\citenamefont {Abbott}\ \emph {et~al.}(2023{\natexlab{b}})\citenamefont {Abbott} \emph {et~al.}}]{KAGRA:2021duu}%
  \BibitemOpen
  \bibfield  {author} {\bibinfo {author} {\bibfnamefont {R.}~\bibnamefont {Abbott}} \emph {et~al.} (\bibinfo {collaboration} {KAGRA, VIRGO, LIGO Scientific}),\ }\bibfield  {title} {\bibinfo {title} {{Population of Merging Compact Binaries Inferred Using Gravitational Waves through GWTC-3}},\ }\href {https://doi.org/10.1103/PhysRevX.13.011048} {\bibfield  {journal} {\bibinfo  {journal} {Phys. Rev. X}\ }\textbf {\bibinfo {volume} {13}},\ \bibinfo {pages} {011048} (\bibinfo {year} {2023}{\natexlab{b}})},\ \Eprint {https://arxiv.org/abs/2111.03634} {arXiv:2111.03634 [astro-ph.HE]} \BibitemShut {NoStop}%
\bibitem [{\citenamefont {Madau}\ and\ \citenamefont {Dickinson}(2014)}]{Madau:2014bja}%
  \BibitemOpen
  \bibfield  {author} {\bibinfo {author} {\bibfnamefont {P.}~\bibnamefont {Madau}}\ and\ \bibinfo {author} {\bibfnamefont {M.}~\bibnamefont {Dickinson}},\ }\bibfield  {title} {\bibinfo {title} {{Cosmic Star Formation History}},\ }\href {https://doi.org/10.1146/annurev-astro-081811-125615} {\bibfield  {journal} {\bibinfo  {journal} {Ann. Rev. Astron. Astrophys.}\ }\textbf {\bibinfo {volume} {52}},\ \bibinfo {pages} {415} (\bibinfo {year} {2014})},\ \Eprint {https://arxiv.org/abs/1403.0007} {arXiv:1403.0007 [astro-ph.CO]} \BibitemShut {NoStop}%
\bibitem [{\citenamefont {Dietrich}\ \emph {et~al.}(2019)\citenamefont {Dietrich}, \citenamefont {Samajdar}, \citenamefont {Khan}, \citenamefont {Johnson-McDaniel}, \citenamefont {Dudi},\ and\ \citenamefont {Tichy}}]{Dietrich:2019kaq}%
  \BibitemOpen
  \bibfield  {author} {\bibinfo {author} {\bibfnamefont {T.}~\bibnamefont {Dietrich}}, \bibinfo {author} {\bibfnamefont {A.}~\bibnamefont {Samajdar}}, \bibinfo {author} {\bibfnamefont {S.}~\bibnamefont {Khan}}, \bibinfo {author} {\bibfnamefont {N.~K.}\ \bibnamefont {Johnson-McDaniel}}, \bibinfo {author} {\bibfnamefont {R.}~\bibnamefont {Dudi}},\ and\ \bibinfo {author} {\bibfnamefont {W.}~\bibnamefont {Tichy}},\ }\bibfield  {title} {\bibinfo {title} {{Improving the NRTidal model for binary neutron star systems}},\ }\href {https://doi.org/10.1103/PhysRevD.100.044003} {\bibfield  {journal} {\bibinfo  {journal} {Phys. Rev. D}\ }\textbf {\bibinfo {volume} {100}},\ \bibinfo {pages} {044003} (\bibinfo {year} {2019})},\ \Eprint {https://arxiv.org/abs/1905.06011} {arXiv:1905.06011 [gr-qc]} \BibitemShut {NoStop}%
\bibitem [{\citenamefont {Pratten}\ \emph {et~al.}(2021)\citenamefont {Pratten} \emph {et~al.}}]{Pratten:2020ceb}%
  \BibitemOpen
  \bibfield  {author} {\bibinfo {author} {\bibfnamefont {G.}~\bibnamefont {Pratten}} \emph {et~al.},\ }\bibfield  {title} {\bibinfo {title} {{Computationally efficient models for the dominant and subdominant harmonic modes of precessing binary black holes}},\ }\href {https://doi.org/10.1103/PhysRevD.103.104056} {\bibfield  {journal} {\bibinfo  {journal} {Phys. Rev. D}\ }\textbf {\bibinfo {volume} {103}},\ \bibinfo {pages} {104056} (\bibinfo {year} {2021})},\ \Eprint {https://arxiv.org/abs/2004.06503} {arXiv:2004.06503 [gr-qc]} \BibitemShut {NoStop}%
\bibitem [{\citenamefont {Usman}\ \emph {et~al.}(2016)\citenamefont {Usman} \emph {et~al.}}]{Usman:2015kfa}%
  \BibitemOpen
  \bibfield  {author} {\bibinfo {author} {\bibfnamefont {S.~A.}\ \bibnamefont {Usman}} \emph {et~al.},\ }\bibfield  {title} {\bibinfo {title} {{The PyCBC search for gravitational waves from compact binary coalescence}},\ }\href {https://doi.org/10.1088/0264-9381/33/21/215004} {\bibfield  {journal} {\bibinfo  {journal} {Class. Quant. Grav.}\ }\textbf {\bibinfo {volume} {33}},\ \bibinfo {pages} {215004} (\bibinfo {year} {2016})},\ \Eprint {https://arxiv.org/abs/1508.02357} {arXiv:1508.02357 [gr-qc]} \BibitemShut {NoStop}%
\bibitem [{\citenamefont {Borhanian}(2021)}]{Borhanian:2020ypi}%
  \BibitemOpen
  \bibfield  {author} {\bibinfo {author} {\bibfnamefont {S.}~\bibnamefont {Borhanian}},\ }\bibfield  {title} {\bibinfo {title} {{GWBENCH: a novel Fisher information package for gravitational-wave benchmarking}},\ }\href {https://doi.org/10.1088/1361-6382/ac1618} {\bibfield  {journal} {\bibinfo  {journal} {Class. Quant. Grav.}\ }\textbf {\bibinfo {volume} {38}},\ \bibinfo {pages} {175014} (\bibinfo {year} {2021})},\ \Eprint {https://arxiv.org/abs/2010.15202} {arXiv:2010.15202 [gr-qc]} \BibitemShut {NoStop}%
\bibitem [{\citenamefont {Gupta}(2024)}]{Gupta:2024bqn}%
  \BibitemOpen
  \bibfield  {author} {\bibinfo {author} {\bibfnamefont {I.}~\bibnamefont {Gupta}},\ }\bibfield  {title} {\bibinfo {title} {{Inferring Small Neutron Star Spins with Neutron Star\textendash{}Black Hole Mergers}},\ }\href {https://doi.org/10.3847/1538-4357/ad49a0} {\bibfield  {journal} {\bibinfo  {journal} {Astrophys. J.}\ }\textbf {\bibinfo {volume} {970}},\ \bibinfo {pages} {12} (\bibinfo {year} {2024})},\ \Eprint {https://arxiv.org/abs/2402.07075} {arXiv:2402.07075 [astro-ph.HE]} \BibitemShut {NoStop}%
\bibitem [{\citenamefont {{Islam}}\ \emph {et~al.}(2023)\citenamefont {{Islam}}, \citenamefont {{Vajpeyi}}, \citenamefont {{Shaik}}, \citenamefont {{Haster}}, \citenamefont {{Varma}}, \citenamefont {{Field}}, \citenamefont {{Lange}}, \citenamefont {{O'Shaughnessy}},\ and\ \citenamefont {{Smith}}}]{Islam:2023zzj}%
  \BibitemOpen
  \bibfield  {author} {\bibinfo {author} {\bibfnamefont {T.}~\bibnamefont {{Islam}}}, \bibinfo {author} {\bibfnamefont {A.}~\bibnamefont {{Vajpeyi}}}, \bibinfo {author} {\bibfnamefont {F.~H.}\ \bibnamefont {{Shaik}}}, \bibinfo {author} {\bibfnamefont {C.-J.}\ \bibnamefont {{Haster}}}, \bibinfo {author} {\bibfnamefont {V.}~\bibnamefont {{Varma}}}, \bibinfo {author} {\bibfnamefont {S.~E.}\ \bibnamefont {{Field}}}, \bibinfo {author} {\bibfnamefont {J.}~\bibnamefont {{Lange}}}, \bibinfo {author} {\bibfnamefont {R.}~\bibnamefont {{O'Shaughnessy}}},\ and\ \bibinfo {author} {\bibfnamefont {R.}~\bibnamefont {{Smith}}},\ }\bibfield  {title} {\bibinfo {title} {{Analysis of GWTC-3 with fully precessing numerical relativity surrogate models}},\ }\href {https://doi.org/10.48550/arXiv.2309.14473} {\bibfield  {journal} {\bibinfo  {journal} {arXiv e-prints}\ ,\ \bibinfo {eid} {arXiv:2309.14473}} (\bibinfo {year} {2023})},\ \Eprint {https://arxiv.org/abs/2309.14473} {arXiv:2309.14473 [gr-qc]} \BibitemShut {NoStop}%
\bibitem [{\citenamefont {Varma}\ \emph {et~al.}(2019)\citenamefont {Varma}, \citenamefont {Field}, \citenamefont {Scheel}, \citenamefont {Blackman}, \citenamefont {Gerosa}, \citenamefont {Stein}, \citenamefont {Kidder},\ and\ \citenamefont {Pfeiffer}}]{Varma:2019csw}%
  \BibitemOpen
  \bibfield  {author} {\bibinfo {author} {\bibfnamefont {V.}~\bibnamefont {Varma}}, \bibinfo {author} {\bibfnamefont {S.~E.}\ \bibnamefont {Field}}, \bibinfo {author} {\bibfnamefont {M.~A.}\ \bibnamefont {Scheel}}, \bibinfo {author} {\bibfnamefont {J.}~\bibnamefont {Blackman}}, \bibinfo {author} {\bibfnamefont {D.}~\bibnamefont {Gerosa}}, \bibinfo {author} {\bibfnamefont {L.~C.}\ \bibnamefont {Stein}}, \bibinfo {author} {\bibfnamefont {L.~E.}\ \bibnamefont {Kidder}},\ and\ \bibinfo {author} {\bibfnamefont {H.~P.}\ \bibnamefont {Pfeiffer}},\ }\bibfield  {title} {\bibinfo {title} {{Surrogate models for precessing binary black hole simulations with unequal masses}},\ }\href {https://doi.org/10.1103/PhysRevResearch.1.033015} {\bibfield  {journal} {\bibinfo  {journal} {Phys. Rev. Research.}\ }\textbf {\bibinfo {volume} {1}},\ \bibinfo {pages} {033015} (\bibinfo {year} {2019})},\ \Eprint {https://arxiv.org/abs/1905.09300} {arXiv:1905.09300 [gr-qc]} \BibitemShut {NoStop}%
\bibitem [{\citenamefont {Boyle}\ \emph {et~al.}(2019)\citenamefont {Boyle} \emph {et~al.}}]{Boyle:2019kee}%
  \BibitemOpen
  \bibfield  {author} {\bibinfo {author} {\bibfnamefont {M.}~\bibnamefont {Boyle}} \emph {et~al.},\ }\bibfield  {title} {\bibinfo {title} {{The SXS Collaboration catalog of binary black hole simulations}},\ }\href {https://doi.org/10.1088/1361-6382/ab34e2} {\bibfield  {journal} {\bibinfo  {journal} {Class. Quant. Grav.}\ }\textbf {\bibinfo {volume} {36}},\ \bibinfo {pages} {195006} (\bibinfo {year} {2019})},\ \Eprint {https://arxiv.org/abs/1904.04831} {arXiv:1904.04831 [gr-qc]} \BibitemShut {NoStop}%
\bibitem [{\citenamefont {Abbott}\ \emph {et~al.}(2021)\citenamefont {Abbott} \emph {et~al.}}]{LIGOScientific:2020ibl}%
  \BibitemOpen
  \bibfield  {author} {\bibinfo {author} {\bibfnamefont {R.}~\bibnamefont {Abbott}} \emph {et~al.} (\bibinfo {collaboration} {LIGO Scientific, Virgo}),\ }\bibfield  {title} {\bibinfo {title} {{GWTC-2: Compact Binary Coalescences Observed by LIGO and Virgo During the First Half of the Third Observing Run}},\ }\href {https://doi.org/10.1103/PhysRevX.11.021053} {\bibfield  {journal} {\bibinfo  {journal} {Phys. Rev. X}\ }\textbf {\bibinfo {volume} {11}},\ \bibinfo {pages} {021053} (\bibinfo {year} {2021})},\ \Eprint {https://arxiv.org/abs/2010.14527} {arXiv:2010.14527 [gr-qc]} \BibitemShut {NoStop}%
\bibitem [{\citenamefont {Romero-Shaw}\ \emph {et~al.}(2020)\citenamefont {Romero-Shaw} \emph {et~al.}}]{Romero-Shaw:2020owr}%
  \BibitemOpen
  \bibfield  {author} {\bibinfo {author} {\bibfnamefont {I.~M.}\ \bibnamefont {Romero-Shaw}} \emph {et~al.},\ }\bibfield  {title} {\bibinfo {title} {{Bayesian inference for compact binary coalescences with bilby: validation and application to the first LIGO\textendash{}Virgo gravitational-wave transient catalogue}},\ }\href {https://doi.org/10.1093/mnras/staa2850} {\bibfield  {journal} {\bibinfo  {journal} {Mon. Not. Roy. Astron. Soc.}\ }\textbf {\bibinfo {volume} {499}},\ \bibinfo {pages} {3295} (\bibinfo {year} {2020})},\ \Eprint {https://arxiv.org/abs/2006.00714} {arXiv:2006.00714 [astro-ph.IM]} \BibitemShut {NoStop}%
\bibitem [{\citenamefont {Speagle}(2020)}]{Speagle:2019ivv}%
  \BibitemOpen
  \bibfield  {author} {\bibinfo {author} {\bibfnamefont {J.~S.}\ \bibnamefont {Speagle}},\ }\bibfield  {title} {\bibinfo {title} {{dynesty: a dynamic nested sampling package for estimating Bayesian posteriors and evidences}},\ }\href {https://doi.org/10.1093/mnras/staa278} {\bibfield  {journal} {\bibinfo  {journal} {Mon. Not. Roy. Astron. Soc.}\ }\textbf {\bibinfo {volume} {493}},\ \bibinfo {pages} {3132} (\bibinfo {year} {2020})},\ \Eprint {https://arxiv.org/abs/1904.02180} {arXiv:1904.02180 [astro-ph.IM]} \BibitemShut {NoStop}%
\bibitem [{\citenamefont {Racine}(2008)}]{Racine:2008qv}%
  \BibitemOpen
  \bibfield  {author} {\bibinfo {author} {\bibfnamefont {E.}~\bibnamefont {Racine}},\ }\bibfield  {title} {\bibinfo {title} {{Analysis of spin precession in binary black hole systems including quadrupole-monopole interaction}},\ }\href {https://doi.org/10.1103/PhysRevD.78.044021} {\bibfield  {journal} {\bibinfo  {journal} {Phys. Rev. D}\ }\textbf {\bibinfo {volume} {78}},\ \bibinfo {pages} {044021} (\bibinfo {year} {2008})},\ \Eprint {https://arxiv.org/abs/0803.1820} {arXiv:0803.1820 [gr-qc]} \BibitemShut {NoStop}%
\bibitem [{\citenamefont {Ajith}\ \emph {et~al.}(2011)\citenamefont {Ajith} \emph {et~al.}}]{Ajith:2009bn}%
  \BibitemOpen
  \bibfield  {author} {\bibinfo {author} {\bibfnamefont {P.}~\bibnamefont {Ajith}} \emph {et~al.},\ }\bibfield  {title} {\bibinfo {title} {{Inspiral-merger-ringdown waveforms for black-hole binaries with non-precessing spins}},\ }\href {https://doi.org/10.1103/PhysRevLett.106.241101} {\bibfield  {journal} {\bibinfo  {journal} {Phys. Rev. Lett.}\ }\textbf {\bibinfo {volume} {106}},\ \bibinfo {pages} {241101} (\bibinfo {year} {2011})},\ \Eprint {https://arxiv.org/abs/0909.2867} {arXiv:0909.2867 [gr-qc]} \BibitemShut {NoStop}%
\bibitem [{\citenamefont {Santamaria}\ \emph {et~al.}(2010)\citenamefont {Santamaria} \emph {et~al.}}]{Santamaria:2010yb}%
  \BibitemOpen
  \bibfield  {author} {\bibinfo {author} {\bibfnamefont {L.}~\bibnamefont {Santamaria}} \emph {et~al.},\ }\bibfield  {title} {\bibinfo {title} {{Matching post-Newtonian and numerical relativity waveforms: systematic errors and a new phenomenological model for non-precessing black hole binaries}},\ }\href {https://doi.org/10.1103/PhysRevD.82.064016} {\bibfield  {journal} {\bibinfo  {journal} {Phys. Rev. D}\ }\textbf {\bibinfo {volume} {82}},\ \bibinfo {pages} {064016} (\bibinfo {year} {2010})},\ \Eprint {https://arxiv.org/abs/1005.3306} {arXiv:1005.3306 [gr-qc]} \BibitemShut {NoStop}%
\bibitem [{\citenamefont {Graham}\ \emph {et~al.}(2023)\citenamefont {Graham} \emph {et~al.}}]{Graham:2022xxu}%
  \BibitemOpen
  \bibfield  {author} {\bibinfo {author} {\bibfnamefont {M.~J.}\ \bibnamefont {Graham}} \emph {et~al.},\ }\bibfield  {title} {\bibinfo {title} {{A Light in the Dark: Searching for Electromagnetic Counterparts to Black Hole\textendash{}Black Hole Mergers in LIGO/Virgo O3 with the Zwicky Transient Facility}},\ }\href {https://doi.org/10.3847/1538-4357/aca480} {\bibfield  {journal} {\bibinfo  {journal} {Astrophys. J.}\ }\textbf {\bibinfo {volume} {942}},\ \bibinfo {pages} {99} (\bibinfo {year} {2023})},\ \Eprint {https://arxiv.org/abs/2209.13004} {arXiv:2209.13004 [astro-ph.HE]} \BibitemShut {NoStop}%
\bibitem [{\citenamefont {Chen}\ \emph {et~al.}(2021)\citenamefont {Chen}, \citenamefont {Cowperthwaite}, \citenamefont {Metzger},\ and\ \citenamefont {Berger}}]{Chen:2020zoq}%
  \BibitemOpen
  \bibfield  {author} {\bibinfo {author} {\bibfnamefont {H.-Y.}\ \bibnamefont {Chen}}, \bibinfo {author} {\bibfnamefont {P.~S.}\ \bibnamefont {Cowperthwaite}}, \bibinfo {author} {\bibfnamefont {B.~D.}\ \bibnamefont {Metzger}},\ and\ \bibinfo {author} {\bibfnamefont {E.}~\bibnamefont {Berger}},\ }\bibfield  {title} {\bibinfo {title} {{A Program for Multimessenger Standard Siren Cosmology in the Era of LIGO A+, Rubin Observatory, and Beyond}},\ }\href {https://doi.org/10.3847/2041-8213/abdab0} {\bibfield  {journal} {\bibinfo  {journal} {Astrophys. J. Lett.}\ }\textbf {\bibinfo {volume} {908}},\ \bibinfo {pages} {L4} (\bibinfo {year} {2021})},\ \Eprint {https://arxiv.org/abs/2011.01211} {arXiv:2011.01211 [astro-ph.CO]} \BibitemShut {NoStop}%
\bibitem [{\citenamefont {Gupta}(2023)}]{Gupta:2022fwd}%
  \BibitemOpen
  \bibfield  {author} {\bibinfo {author} {\bibfnamefont {I.}~\bibnamefont {Gupta}},\ }\bibfield  {title} {\bibinfo {title} {{Using grey sirens to resolve the Hubble\textendash{}Lema\^\i{}tre tension}},\ }\href {https://doi.org/10.1093/mnras/stad2115} {\bibfield  {journal} {\bibinfo  {journal} {Mon. Not. Roy. Astron. Soc.}\ }\textbf {\bibinfo {volume} {524}},\ \bibinfo {pages} {3537} (\bibinfo {year} {2023})},\ \Eprint {https://arxiv.org/abs/2212.00163} {arXiv:2212.00163 [gr-qc]} \BibitemShut {NoStop}%
\bibitem [{\citenamefont {Chen}\ \emph {et~al.}(2024)\citenamefont {Chen}, \citenamefont {Ezquiaga},\ and\ \citenamefont {Gupta}}]{Chen:2024gdn}%
  \BibitemOpen
  \bibfield  {author} {\bibinfo {author} {\bibfnamefont {H.-Y.}\ \bibnamefont {Chen}}, \bibinfo {author} {\bibfnamefont {J.~M.}\ \bibnamefont {Ezquiaga}},\ and\ \bibinfo {author} {\bibfnamefont {I.}~\bibnamefont {Gupta}},\ }\bibfield  {title} {\bibinfo {title} {{Cosmography with next-generation gravitational wave detectors}},\ }\href {https://doi.org/10.1088/1361-6382/ad424f} {\bibfield  {journal} {\bibinfo  {journal} {Class. Quant. Grav.}\ }\textbf {\bibinfo {volume} {41}},\ \bibinfo {pages} {125004} (\bibinfo {year} {2024})},\ \Eprint {https://arxiv.org/abs/2402.03120} {arXiv:2402.03120 [gr-qc]} \BibitemShut {NoStop}%
\end{thebibliography}%

\end{document}